%% file: main.tex
\documentclass[sigconf]{acmart}

\usepackage{adjustbox}
\usepackage[super]{nth}
\usepackage{amsmath}
\usepackage{amsthm}
\usepackage{mathtools}
\usepackage{bm}
\usepackage{dsfont}
\usepackage{graphicx}
\usepackage{caption}
\usepackage{subcaption}
\usepackage{placeins}
\usepackage{float}
\usepackage{enumitem}
\usepackage{array}
\usepackage{makecell}
\usepackage{multirow}
\usepackage{tabularx}
\usepackage{tcolorbox}
\usepackage{xcolor}

\DeclareMathOperator*{\argmin}{arg\,min}
\usepackage{multicol}
\usepackage{lipsum} 

\usepackage{hyperref}
\hypersetup{
    colorlinks=true,
    linkcolor=blue,
    filecolor=magenta,      
    urlcolor=cyan,
    pdftitle={Overleaf Example},
    pdfpagemode=FullScreen,
}

\theoremstyle{definition}

\usepackage{xcolor}

\newtoggle{comments}
 \toggletrue{comments} 
\NewEnviron{doweprint}[1]{%
  \iftoggle{#1}{\BODY}{}%
}

\newcounter{todocounter}
\usepackage{xcolor,pifont}
\definecolor{cycle2}{RGB}{106, 191, 0}
\definecolor{cycle3}{RGB}{191, 0, 0}

\begin{document}
\title{Building Transparency in Deep Learning-Powered Network Traffic Classification: A Traffic-Explainer Framework}

\author{Riya Ponraj}
\affiliation{%
  \institution{University of Oregon}
  \country{}
}

\author{Ram Durairajan}
\affiliation{%
  \institution{University of Oregon,  Link Oregon}
  \country{}
}

\author{Yu Wang}
\affiliation{%
  \institution{University of Oregon}
  \country{}
}

\begin{abstract}
Recent advancements in deep learning (DL) have significantly enhanced the performance and efficiency of traffic classification in networking systems. However, the lack of transparency in their predictions and decision-making has made network operators reluctant to deploy DL-based solutions in production networks. 
To tackle this challenge, we propose \texttt{Traffic-Explainer}, a model-agnostic and input-perturbation-based traffic explanation framework. By maximizing the mutual information between predictions on original traffic sequences and their masked counterparts, Traffic-Explainer automatically uncovers the most influential features driving model predictions. Extensive experiments demonstrate that Traffic-Explainer improves upon existing explanation methods by approximately 42\%.
Practically, we further apply Traffic-Explainer to identify influential features and demonstrate its enhanced transparency across three critical tasks: application classification, traffic localization, and network cartography.
For the first two tasks, Traffic-Explainer identifies the most decisive bytes that drive predicted traffic applications and locations, uncovering potential vulnerabilities and privacy concerns. In network cartography, Traffic-Explainer identifies submarine cables that drive the mapping of traceroute to physical path, enabling a traceroute-informed risk analysis.
Our implementation is publicly available at \url{https://anonymous.4open.science/r/TrafficExplainer-5E2E/README.md}.

\end{abstract}

\maketitle

\input{introduction}

\input{RelatedWork}

\input{Preliminary}

\input{Framework}

\input{Experiment}
\input{Conclusion}


\bibliographystyle{ACM-Reference-Format}
\bibliography{reference}

\appendix

\input{Appendix}

\end{document}

%% file: introduction.tex
\begin{figure}[t!]
\centering
\includegraphics[width=0.485\textwidth]{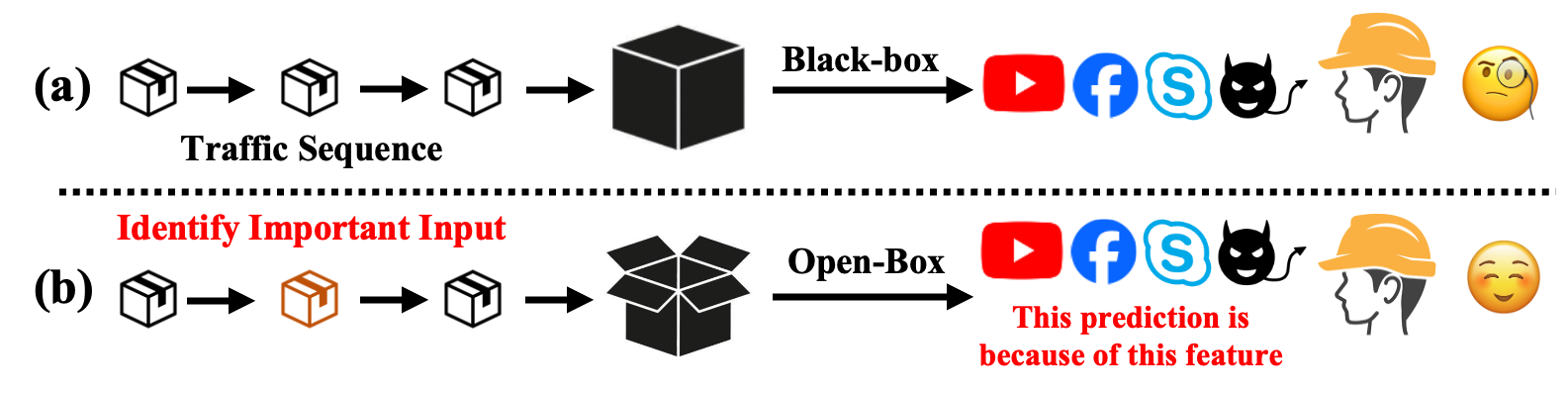}
\vspace{-4ex}
\caption{(a) \textbf{Untransparent scenario:} 
DL solutions that solely present predictions with no explanation lose the trust of network operators. (b) \textbf{Transparent scenario:} 
DL solutions that offer explanations for their decision-making gain by highlighting the critical unit driving the prediction of the traffic input sequence, earning the trust of network operators.}
\label{fig-motivation}
\vspace{-4ex}
\end{figure}

\section{Introduction}
Network traffic classification~\cite{zhang2023tfe, lin2022bert}, which infers properties of network transmissions from packet flows between interconnected devices in networked systems, supports numerous critical tasks, such as application classification~\cite{zhang2023tfe, lin2022bert}, traffic localization~\cite{qian2024netbench}, and network cartography~\cite{ram@nautilus, durairajan2015intertubes}. Recent advances in deep learning (DL) have led to powerful automated solutions for traffic classification. However, decisions made by DL models are based on learned features and lack explainability. Although earlier feature-engineering-based traffic classifiers, such as fingerprint matching~\cite{taylor2017robust, van2020flowprint}, are inherently explainable by manually extracting features (e.g., device and certificate) based on domain expertise, they are less effective and nonadaptive. DL-based models like ET-Bert~\cite{lin2022bert} and TFE-GNN~\cite{zhang2023tfe} operate in latent embedding spaces, where internal representations (i.e., hidden neurons) have no clear correspondence with human-understandable concepts or predicted labels. This opacity limits transparency and has made network operators hesitant to adopt DL-based solutions in traffic classification~\cite{knofczynski2022arise, elfandibootstrapping}. In high-stakes environments, such as government networks or critical infrastructure, a DL model may flag traffic as suspicious or associated with prohibited applications (e.g., BitTorrent) without revealing whether the decision stemmed from payload patterns, header anomalies, or timing irregularities. Without such insight, network operators cannot verify actions like blocking or rerouting, complicating compliance and auditability.

To improve the transparency, conventional DL-based explanation methods~\cite{adadi2018peeking, verma2020counterfactual} can be naturally employed. 
For example, gradient-based methods such as Saliency Maps and Grad-CAM~\cite{simonyan2013deep, selvaraju2017grad, wang2024gradient} can be applied to uncover the most determinant factors based on the input gradient.
Despite these well-established explanation methods in the general DL literature~\cite{ribeiro2016should, lundberg2017unified, ying2019gnnexplainer, amann2020explainability, zhao2024explainability}, their adoption in traffic classification or even broader networking system domains remains limited. Although traditional rule-based approaches are self-explainable~\cite{knofczynski2022arise}, their reliance on hand-crafted features makes them inflexible and rigid to generalize across different datasets or tasks. To date, only a few works have explicitly bridged explainability and networking systems. NetXplain~\cite{pujol2021netxplain} focuses solely on explaining traffic delays, and Hybrid Explainability~\cite{elfandibootstrapping} is restricted to post-hoc interpretation of non-DL models such as decision trees. There is still no unified framework that systematically explains the behavior of DL-based traffic classification models.

Given the criticality yet the nascent state of building the transparency of DL-powered traffic classification, this paper presents an explanation framework, \texttt{Traffic-Explainer}, designed explicitly for DL-based traffic classification applications. Achieving this goal requires addressing three key challenges. \textit{(1) First, we need to determine the explanation object that balances specificity (i.e., encoding sufficient class-desired information to enable the explanation) and generalizability (i.e., being applicable across various traffic classification scenarios).} Since many network traffic classification problems can be formulated as sequential classification (e.g., application classification based on sequential traffic packets and network cartography based on sequential round-trip times (RTTs)), Traffic-Explainer treats each sequence as the basic explanation object and aims to identify the most critical units within the sequence responsible for its classification. For instance, in traffic application classification, where sequences consist of bytes or packets, explanations should highlight the critical bytes and byte-byte interactions that determine the classification outcome. In network cartography, where traffic sequences consist of sub-segments annotated with RTTs across hops, effective explanations should pinpoint the specific RTT hop that best characterizes the underlying physical fiber cable carrying that particular traffic. \textit{(2) Second, the underlying classifier should be capable of handling sequential data, achieving high performance, and remaining accessible to practitioners.} To this end, we apply our proposed Traffic-Explainer to explain predictions from two widely-used DL architectures: transformer, for its advanced capabilities on sequential data, and multi-layer perceptron (MLP), for its simplicity and usability. This design choice underscores the model-agnostic nature of our Traffic-Explainer. \textit{(3) Thirdly, the generated explanations should be naturally interpretable to domain experts such as network operators.} To achieve this, our Traffic-Explainer formulates explanation generation as a mutual information maximization~\cite{ying2019gnnexplainer}, aiming to identify the most informative input units (e.g., bytes and RTTs) that maximize interpretability from the network operator perspective. We summarize contributions as follows:

\begin{itemize}[leftmargin=*]
\item \textbf{Novel Explanation Problem on Traffic Classification:} This work pioneers the explanation of DL-based traffic classification, aiming to enhance the transparency of DL-driven decisions and provide domain insights to assist network operators.

\item \textbf{Systematic Explanation Framework for Traffic Classification:} We introduce Traffic-Explainer, a model-agnostic and input-perturbation-based explanation framework that identifies the most influential features by maximizing the mutual information between predictions on original inputs and their masked versions. Extensive experiments validate its effectiveness, efficiency, and transferability of explanations across DL models.

\item \textbf{Three Real-world Applications of Traffic Classification:} We demonstrate the efficacy of Traffic-Explainer through three representative use cases. For application classification and traffic localization, Traffic-Explainer identifies the most characteristic bytes driving instance-level predictions, such as a given application or location. For network cartography, the generated explanations enable network operators to automatically pinpoint the most likely submarine cables traversed by a given traffic traceroute, resulting in more transparent logical-to-physical dependencies.

\end{itemize}



%% file: RelatedWork.tex
\section{Related Work}\label{sec-relatedwork}

\textbf{Explainable DL-based Network Traffic Classification.}
The ubiquity of network traffic coupled with a shortage of skilled operators has driven the adoption of deep learning for automated traffic classification~\cite{knofczynski2022arise, elfandibootstrapping, kaloudi2020ai, abdullahi2022detecting}. However, their black-box nature raises concerns about transparency and motivates the development of explanation techniques.
Explanation methods can be generally categorized into four types: gradient, perturbation, surrogate, and decomposition methods. Gradient methods analyze the gradients of model outputs with respect to input features~\cite{selvaraju2020grad}. Perturbation-based methods modify input data and observe changes in model outputs~\cite{robnik2018perturbation}. Surrogate methods, such as LIME~\cite{ribeiro2016should}, approximate model behavior using simpler interpretable models. Decomposition methods break down predictions into additive contributions of input features~\cite{feng2023degree}. Despite their broad application in general DL~\cite{ribeiro2016should, lundberg2017unified, ying2019gnnexplainer, amann2020explainability, zhao2024explainability}, 
they have seen limited adoption in DL-powered solutions~\cite{elfandibootstrapping}. For instance, rule-based traffic classification is inherently explainable~\cite{knofczynski2022arise}. However, its static handcrafted logic lacks adaptability across diverse datasets and tasks. Notably, two ML-based networking explanation frameworks, such as NetXplain~\cite{pujol2021netxplain} and Hybrid Explainability~\cite{elfandibootstrapping}, remain narrow in scope, where the former targets only traffic delay explanations, while the latter focuses on post-hoc interpretations for non-DL models. To date, no framework systematically explains DL behaviors of traffic classification. This gap motivates us to propose Traffic-Explainer, which adapts a perturbation-based masking strategy to identify key sequence units for explanation. We next review its three applications.


\textbf{Application Classification and Traffic Localization.}
Traffic classification aims to categorize sequential traffic flows composed of byte sequences and has been used in application classification and traffic localization~\cite{azab2024network, qian2024netbench, lin2022bert, zhang2023tfe}. Traditional methods rely on handcrafted features, such as traffic fingerprints and statistics, which require extensive domain expertise and are limited to specific traffic scenarios~\cite{sirinam2018deep, liu2019fs}. Recent DL approaches (e.g., Transformers and Graph Neural Networks) to automatically learn representations for traffic classification~\cite{lin2022bert, zhang2023tfe}. However, the black-box nature of traffic classification models makes their predictions difficult to interpret, highlighting the need for explanation techniques to identify the key components in each traffic sequence that drive instance-level predictions and characterize class-level signatures. For example, these explanation techniques are expected to reveal which byte patterns are most influential in distinguishing the "Chat" application from "P2P", or distinguishing locations where traffic occurs.

\textbf{Network Cartography.}
Network cartography aims to establish the dependency between the logical traffic and the physical infrastructure layer in networking systems~\cite{dan2021ip}. Given the routing of traffic along IP-level paths, the goal is to infer the physical paths (e.g., the terrestrial or submarine fiber-optic cables) traversed by the traffic. Prior efforts~\cite{ram@nautilus, anderson@igdb} have relied on commonsense heuristics, such as speed-of-light filters, geographic proximity, and networking domain expertise. However, these signals are noisy, incomplete, and hard-coded~\cite{thiagarajan2025aleph}. For instance, cables owned by the same provider can be challenging to differentiate using ownership data alone, and IP geo-locating services are notoriously known to produce erroneous results. To address these limitations, recent work~\cite{ramanathan2025leveraging} has explored temporal patterns within traceroutes (i.e., the sequence of RTTs) to infer physical cable mappings. Building on this direction, our proposed Traffic-Explainer can be used as a post-hoc analysis tool to identify the most discriminative temporal features, such as a specific hop with a characteristic RTT, that reveal the property of the underlying physical cable. These features can serve as cable signatures, supporting risk assessment/resilience planning in networking design~\cite{durairajan2015intertubes, sanchez2014inter}.

%% file: Preliminary.tex
\section{Preliminary}
Since many traffic classification problems, such as application classification, traffic localization, and network cartography in Section~\ref{sec-relatedwork}, operate on sequence-based traffic data, we adopt the sequence as the fundamental unit for DL-based traffic classification and the subsequent explanation. We now introduce notations and problems.

\subsection{Mathematical Notations}
Let \(\mathcal{X} = \{(\mathcal{X}^i, \mathcal{Y}^i)\}_{i = 1}^N\) represent a set of \(N\) traffic sequences, where the \(i^{\text{th}}\) sequence consists of a series of units \(\mathcal{X}^i = \{\mathcal{X}^{i}_{j}\}_{j = 1}^{|\mathcal{X}^i|}\) and its corresponding one-hot label \(\mathcal{Y}^{i, *} \in \{0,1\}^{C}\) with $C$ being the class number. In the context of application classification and traffic localization, each sequence \(\mathcal{X}^i\) corresponds to a sequence of bytes, where each byte \(\mathcal{X}_j^i\) takes values in the range \([0, 255]\), and the label \(\mathcal{Y}^{i, *}\) represents the downstream application (e.g., YouTube, Skype, and Facebook) or traffic locations (e.g., US, China, and India). In network cartography, each sequence \(\mathcal{X}^i\) consists of logical-layer measurements, where each measurement \(\mathcal{X}_j^i\) is a real-valued RTT, and the label \(\mathcal{Y}^{i, *}\) corresponds to the traversed physical infrastructure (e.g., fiber links). Assuming the traffic classifier as $g_{\boldsymbol{\Theta}_g}$ (e.g., transformer and MLP) and the proposed Traffic-Explainer as $h_{\boldsymbol{\Theta}_h}$ parameterized respectively with $\boldsymbol{\Theta}_g, \boldsymbol{\Theta}_h$, we formulate the explanation problem of sequence-level classification in the following.


\subsection{Problem Statement}
Given a traffic sequence $\mathcal{X}^i$, we aim to:
\begin{itemize}[leftmargin=*]
    \item \textbf{Learn an optimal classifier $g_{\boldsymbol{\Theta}_g^{*}}$}: $\mathcal{X}^i \rightarrow \mathcal{Y}^i$ so that the predicted class $\mathcal{Y}^i$ matches the ground-truth class  $\mathcal{Y}^{i, *}$.

    \item \textbf{Learn an optimal explainer $h_{\boldsymbol{\Theta}_h^{*}}$}: $(\mathcal{X}^i, \mathcal{Y}^i, g_{\boldsymbol{\Theta}_{g}^{*}}) \rightarrow \mathbf{M}^{i, *}$ so that the learned feature mask $\mathbf{M}^{i, *} \in [0,1]^{|\mathcal{X}^i|}$ could maximally explain the prediction $\mathcal{Y}^i$ by the classifier $g_{\boldsymbol{\Theta}_g^*}$ over $\mathcal{X}^i$.
\end{itemize}

\noindent Notably, $\mathbf{M}^{i, *}$ denotes the feature mask over each unit in the sequence $\mathcal{X}^i$, where each value ranges between 0 and 1 and represents the importance of that unit in contributing to the model prediction. This problem setup can be generalized to many sequence-based traffic classification applications, such as application classification, traffic localization, and network cartography.

%% file: Framework.tex
\begin{figure*}[t!]
    \centering
    \includegraphics[width=1.0\textwidth]{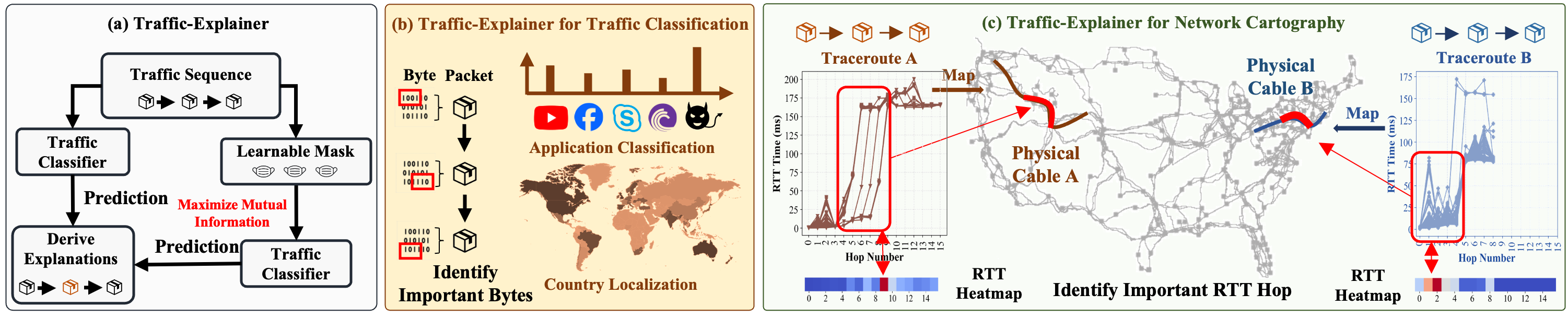}
    \vspace{-2ex}
    \caption{(a) Given a traffic sequence $\mathcal{X}^i = \{\mathcal{X}_j^i\}_{j=1}^{|\mathcal{X}^i|}$ consisting of $|\mathcal{X}^i|$ feature units, an MLP or Transformer-based traffic classifier first predicts the corresponding traffic label.
    To provide interpretability, our proposed Traffic-Explainer identifies the Top-K most influential input units that drive the model prediction by optimizing input masks via mutual information maximization. (b) In application classification and traffic localization, Traffic-Explainer produces explanations in the form of individual bytes in the traffic packet sequence that are most critical for the model’s decision.(c) In the network cartography, Traffic-Explainer highlights the specific RTT-hop in a traceroute sequence that corresponds to the physical submarine cable being traversed, enabling transparent mapping between logical observations and physical infrastructure.}
    \vspace{-2ex}
    \label{fig-framework}
\end{figure*}

\section{Framework}\label{sec-model}
Building on the above notations and problem formulation, we next briefly introduce our transformer-based traffic classifier since understanding its architecture informs the design of our proposed Traffic-Explainer. \textit{Notably, Traffic-Explainer is an explanation model that differs from the classifier itself. Furthermore, Traffic-Explainer directly operates in the input space to identify the most influential features (i.e., units within a sequence), making it inherently model-agnostic.} While our primary classifier is based on the Transformer architecture, we also evaluate Traffic-Explainer using a simpler backbone, such as a multi-layer perceptron (MLP), to demonstrate its generality and adaptability across model types. Given the simplicity and well-known structure of MLPs, we omit a detailed discussion of their architecture and instead focus on briefly introducing the Transformer-based backbone.

\subsection{Transformer-based Traffic Classifier}\label{sec-classifier}
Transformer-based traffic classifier $g_{\boldsymbol{\Theta}_g}$ comprises self-attention, feed-forward, and pooling layers, which are introduced next..

\vspace{-0.5ex}
\subsubsection{Unit Tokenization}
Each sequence $\mathcal{X}^i$ consists of a set of units, where unit $\mathcal{X}_j^i$ is mapped to a $d$-dimensional embedding via a learnable embedding matrix $\mathbf{E} \in \mathbb{R}^{|\mathcal{V}| \times d}$, where $|\mathcal{V}|$ denotes the vocabulary size. For instance, in application classification, each byte $b_k$ in a packet takes a value in $[0, 256]$, where 256 serves as a padding token to standardize packet lengths. We retrieve the embedding of the $j^\text{th}$ unit as $\mathbf{e}^i_{j} = \mathbf{E}[\mathcal{X}_j^i, :],~~\forall~\mathcal{X}_j^i \in \mathcal{X}^i$. For real-valued RTT sequences in network cartography, the units are projected into a continuous embedding space via a learnable linear transformation.

\vspace{-0.5ex}
\subsubsection{Iterative Self-Attention and Feed-Forward}
To capture dependencies among units, we apply self-attention. Since ordering information is essential (e.g., byte order in packets or temporal order in RTTs), we incorporate positional encoding $\boldsymbol{\Phi}_j$ into each unit embedding. The transformer then applies self-attention over the sequence to produce contextual embeddings: $\{\mathbf{h}^i_j\}_{j=1}^{|\mathcal{X}^i|} = \text{Self-ATT}(\{(\mathbf{e}^i_j, \boldsymbol{\Phi}_j)\}_{j = 1}^{|\mathcal{X}^i|})$. These embeddings are subsequently passed through a feed-forward layer. This process is repeated iteratively across multiple transformer layers to progressively refine the representation.

\vspace{-0.5ex}
\subsubsection{Pooling and Classification}
After obtaining contextual unit embeddings $\{\mathbf{h}_j^i\}_{j=1}^{|\mathcal{X}^i|}$, we aggregate them into a fixed-size sequence embedding via a pooling: $\mathbf{F}^i = \text{Pooling}(\{\mathbf{h}_j^i\}_{j=1}^{|\mathcal{X}^i|})$.
where the typical pooling operation could be mean-pooling, and the obtained sequence embedding $\mathbf{F}^i$ serves as the final embedding for classification. Given the ground-truth class $\mathcal{Y}^{i, *}$, we optimize the classifier parameters $\boldsymbol{\Theta}_g$ by minimizing the cross-entropy loss: $\boldsymbol{\Theta}_{g}^{*} = \argmin_{\boldsymbol{\Theta}_g} \sum_{i = 1}^{N}\sum_{c = 1}^{C} \mathcal{Y}^{i, *}_{c} \log \mathcal{Y}^{i}_{c}$ where $\mathcal{Y}^{i} \in \mathbb{R}^{C}$ is the predicted class distribution after applying softmax normalization and linear mapping on $\mathbf{F}^i$. This formulation ensures that our transformer-based classifier supports both byte/RTT sequences for application classification, country localization, and network cartography.


\subsection{Traffic-Explainer}\label{sec-explain}
After introducing the traffic classifier, this section focuses on developing our proposed explanation framework, Traffic-Explainer. 
We aim to identify the most important units that are responsible for prediction decisions. Formally, given a sequence $\mathcal{X}^i = \{\mathcal{X}_j^i\}_{j = 1}^{|\mathcal{X}^i|}$, we seek to extract a subset of units $\widehat{\mathcal{X}}^i \subseteq \mathcal{X}^i$ that are most responsible for its prediction $\mathcal{Y}^i$. Since the notion of a fundamental unit varies across applications (e.g., bytes in traffic classification, RTTs in network cartography), Traffic-Explainer provides a generalizable approach by centering explanations on these core sequence components. For instance, bytes form the fundamental building blocks of traffic flows across diverse protocols, while RTTs offer ubiquitous network performance measurements. By identifying the most critical units for classification, our method enhances interpretability across heterogeneous domains. Moreover, our framework extends beyond individual units to capture unit-unit interactions by refining the masking mechanism to operate at the self-attention layer rather than the input level. We demonstrate the explanation effectiveness at the byte level and the byte-byte interaction level in Section~\ref{sec-expr}.

Given a sequence of units $\mathcal{X}^{i}$, to identify the most important units, we initialize a learnable unit masking matrix $\mathbf{M}^i\in [0, 1]^{|\mathcal{X}^i|}$ with $\mathbf{M}^i_{j}$ denoting the importance of unit $j$ in contributing to the model prediction of $\mathcal{X}^i$ to be $\mathcal{Y}^i$. The traffic classifier predicts with the masked sequence, i.e., its first self-attention layer becomes: $\{\widehat{\mathbf{h}}^i_j\}_{j=1}^{|\mathcal{X}^i|} = \text{Self-ATT}(\{(\mathbf{e}^i_{j} * \sigma{(\mathbf{M}^i_{j})}, \boldsymbol{\Phi}_{j})\}_{j = 1}^{|\mathcal{X}^i|})$ where $\sigma$ is the sigmoid function mapping the mask score to a value between 0 and 1. The transformed units $\{\widehat{\mathbf{h}}^i_j\}_{j=1}^{|\mathcal{X}^i|}$ are then aggregated via pooling followed by the linear classifier following Section~\ref{sec-classifier}. A higher value of $\mathbf{M}_{j}^i$ after optimization by explanation objective indicates the higher importance of unit $\mathcal{X}_j^i$ in predicting the sequence label. We next introduce explanation objectives at the local-instance/global-class levels for optimizing the unit-level masking $\mathbf{M}^i$.

\subsubsection{Local Instance Explanation}
For each sequence $\mathcal{X}^i$ with predicted class $\mathcal{Y}^i = g_{\boldsymbol{\Theta}^{*}_g}(\mathcal{X}^i)$, we optimize explanation by maximizing mutual information (MI) between the prediction $\mathcal{Y}^i$ and its explanation $\widehat{\mathcal{X}}^i$ (i.e., a subset of units):
\begin{equation}\label{eq-explain-MI}
    \max_{\widehat{\mathcal{X}}^i}\text{MI}(\mathcal{Y}^i, \widehat{\mathcal{X}}^i) = H(\mathcal{Y}^i) - H(\mathcal{Y}^i|\widehat{\mathcal{X}}^i)
\end{equation}
For traffic sequence $\mathcal{X}^i$, MI quantifies the change in the prediction $\mathcal{Y}^i$ when the sequence $\mathcal{X}^i$ is masked to become the explained sub-sequence $\widehat{\mathcal{X}}^i$. For example, if removing the unit $\widehat{\mathcal{X}}^i_j$ strongly decreases the prediction probability of $\max_{c\in\mathcal{C}}{\mathcal{Y}}^i_c$, the unit $\widehat{\mathcal{X}}^i_j$ is naturally a good counterfactual explanation for the prediction of sequence $\mathcal{X}^i$. 
Maximizing the mutual information between the predicted label distribution $\mathcal{Y}^i$ and explanation (i.e., the masked sequence) $\widehat{\mathcal{X}}^i$ equals minimizing the conditional entropy $H(\mathcal{Y}^i|\widehat{\mathcal{X}}^i)$:
\begin{equation}\label{eq-confidence-explain}
\small
    \mathbf{M}^{i, *} = \argmin_{\mathbf{M}^{i}} H(\mathcal{Y}^i|\widehat{\mathcal{X}}^i) = \argmin_{\mathbf{M}^{i}}-\mathbb{E}_{\mathcal{Y}^i|\mathcal{X}^i}[\log P_{g_{\boldsymbol{\Theta}^{*}_g}}(\widehat{\mathcal{Y}}^i|\widehat{\mathcal{X}}^i)].
\end{equation}
The explanation for prediction $\mathcal{Y}$ is thus a subsequence of units $\mathcal{X}$ that minimizes uncertainty of $g_{\boldsymbol{\Theta}^{*}_g}$, following the intuition that \textit{the explanation should be the ones that if only keep features identified by the explainer and remove all other input features, the model would become more confident about its original prediction distribution.}

Rather than explaining through the model confidence, the network operators sometimes care about \textit{``why does the trained traffic classifier predict a certain class label?".} Therefore, we modify the conditional entropy objective in Eq.~\eqref{eq-confidence-explain} towards the class label $c$:
\begin{equation}\label{eq-instance-explain}
    \mathbf{M}^{i, *}(c) = \argmin_{\mathbf{M}^{i}} -\sum_{j = 1}^{C}\mathds{1}[\mathcal{Y}_j^i = c]\log P_{g_{\boldsymbol{\Theta}_g^{*}}}(\widehat{\mathcal{Y}}^i_j|\widehat{\mathcal{X}}^i).
\end{equation}
The explanation for prediction is thus a subsequence of units $\widehat{\mathcal{X}}$ that, if only keep them and remove all other units in the sequence, would maximize the prediction score of the model towards its original predicted class (as compared to its original prediction distribution as before). We empirically find that this objective slightly outperforms the previous confidence objective in explanation.

\subsubsection{Global Class Explanation}
Both previous explanations are only at the instance level. In real-world applications, what is more interesting is why the model always makes certain predictions about a group of instances, e.g., the ones belonging to the same class. This motivates us to explain the predictions at the global class level. For example, identifying the most important units driving predictions towards a certain class across all instances.
\begin{equation}\label{eq-class-explain}
    \small
    \mathbf{M}^{*} = \argmin_{\mathbf{M}} -\sum_{i = 1}^{N}\sum_{j = 1}^{C}\mathds{1}[\mathcal{Y}^{i}_{j} = c]\log P_{g_{\boldsymbol{\Theta}_g^{*}}}(\widehat{\mathcal{Y}}^{i}_{j}|\widehat{\mathcal{X}}^i)
\end{equation}

\subsubsection{Mask Regularization}
Blindly optimizing the mask $\mathbf{M}$ following the above explanation-based objectives Eq.~\eqref{eq-confidence-explain}-\eqref{eq-class-explain} may lead to trivial explanations, e.g., $\widehat{\mathcal{X}} = \mathcal{X}$, where all units in the original sequence are tagged important, as it would naturally encompass information necessary to explain the model prediction. Moreover, a widely adopted assumption in feature selection and sparse representation learning is that model prediction should primarily be attributed to a subset of the inputs~\cite{li2017feature}. To prevent the trivial use of the entire sequence as the explanation and also consider the sparsity principle, we impose a predefined budget $B$ to limit the magnitude of the explanation mask. We weighted combine the explanation loss $\mathcal{L}^{\text{Explain}}$ and the budget constrain loss:
\begin{equation}
    \small
    \mathbf{M}^{*} = \argmin_{\mathbf{M}} \mathcal{L} = \alpha_1\text{ReLU}(||\mathbf{M}||_1 - B) + \alpha_2\mathcal{L}^{\text{Explain}},
\end{equation}
where $\mathcal{L}^{\text{Explain}}$ could refer to any of the previous three explanation losses defined in Eq~\eqref{eq-confidence-explain}-\eqref{eq-class-explain}. \textit{Intuitively, minimizing $\mathcal{L}$ identifies the most informative units while maximally excluding units irrelevant to the explanation, thereby refining the final explanation.}

Although the presented framework only focuses on identifying the most critical units, it can be easily tailored to identify the most important unit-unit interactions by masking the unit self-attention. Therefore, we also include this level of explanation in Table~\ref{tab-local-instance-res}.

%% file: Experiment.tex
\section{Experiments}\label{sec-expr}
In this section, we conduct experiments to evaluate the effectiveness, efficiency, and transferability of the explanation discovered by Traffic-Explainer. We introduce experimental settings below. 

\subsection{Experimental Settings}

\subsubsection{Application Tasks and Dataset Collection}
We demonstrate the practical usage of our proposed Traffic-Explainer through three application tasks, including their objectives and dataset collection, explanation baselines, evaluation strategies, and hyperparameter settings. Comprehensive details are in Appendix~\ref{app-data}.

\begin{itemize}[leftmargin=*]
    \item \textbf{Application 1 - Application Classification~\cite{lin2022bert, zhang2023tfe}}: This task classifies the application (e.g., YouTube, Skype) based on traffic flows represented as byte sequences. The explanation identifies the most influential bytes that drive the predicted application. Following~\cite{zhang2023tfe}, we validate the proposed Traffic-Explainer on four datasets: ISCX-VPN, ISCX-NonVPN, ISCX-Tor, and ISCX-NonTor. We use SplitCap to obtain bidirectional flows and increase the training samples in ISCX-Tor by dividing each flow into 60-second non-overlapping blocks. 

    \item \textbf{Application 2 - Traffic Localization~\cite{qian2024netbench, wang2024lens}}: aims to identify the country where a given sequence of network traffic occurred. This is conceptually framed similarly to traffic classification, where given a network flow (hex-encoded data), it predicts the country label of its origin or destination. The explanation is to identify the most critical bytes that contribute to the model prediction of the country. Two datasets, IOS-Cross Platform and Android-Cross Platform~\cite{van2020flowprint, qian2024netbench}, are used to showcase Traffic-Explainer in identifying traffic country localization. These two datasets comprise user-generated data for 215 Android and 196 iOS apps in the US, China, and India.

    \item \textbf{Application 3 - Network Cartography~\cite{Ram@mapping}}: focuses on mapping traffic traceroute data to physical submarine cables. Each data instance represents the routing of traffic along IP-level paths, with the goal of inferring the corresponding physical paths (e.g., terrestrial or submarine fiber-optic cables). The explanation task aims to identify the most critical RTT hops that contribute to the mapping of traffic to physical cables. We collect traceroutes from predefined source locations to destinations. Using RIPE Atlas probes~\cite{holterbach@ripe}, we select target countries and direct probes to a set of international servers. Over a three-day collection period, we gathered 5,000 traceroutes across 10 unique source-destination pairs, which serve as classification labels. Among these, three key classes include: (1) Seattle, US to Yokohama, Japan (submarine route); (2) Seattle, US to San Jose, US (terrestrial route); and (3) Virginia Beach, US to San Sebastian, France (submarine route).
\end{itemize}

Details about dataset statistics are presented in Appendix~\ref{app-data}.


\newpage
\subsubsection{Explanation Baselines}
To benchmark the explanation by the proposed Traffic-Explainer, in the first application classification, we compare it with five explanation baselines: \textbf{Random}-we randomly select the Top-K bytes or byte-byte interactions and treat them as the explanation; \textbf{Saliency Map/Gradient-based Methods}~\cite{simonyan2013deep}-we calculate the gradient of the output prediction with respect to either each byte or each byte-byte interaction and then select the ones with top-K gradient magnitude as the explanation; \textbf{Self-Attention}~\cite{hao2021self}-we use the transformer attention as the importance score for each byte-byte interaction and select the top-K ones as the explanation; \textbf{LIME}~\cite{ribeiro2016should}-A local surrogate model is trained by perturbing the input (e.g., masking byte values) and observing the model's output. The learned weights of the surrogate model approximate the importance of each byte or byte value, from which the Top-K most influential ones are selected; \textbf{SHAP}~\cite{lundberg2017unified}-SHAP estimates the contribution of each byte or byte-byte interaction to the model's output using Shapley values from cooperative game theory. The Top-K most important features are selected based on their absolute Shapley values.

\subsubsection{Evaluation Metrics}
Four explanation evaluation metrics are used: Fidelity, Accuracy, Counterfactual Fidelity, and Counterfactual Accuracy~\cite{bodria2023benchmarking, alangari2023exploring}. For each instance, we extract the Top-K most important bytes as the explanation and assess how the model prediction changes when these bytes are either (1) removed for counterfactual fidelity and counterfactual accuracy evaluation or (2) exclusively retained for fidelity and accuracy evaluation. Due to space limitations, we present their comprehensive definition and computation equation in the Appendix~\ref{app-evaluation}.


\subsubsection{Implementation Details and Hyperparameters} We train the traffic classifier using the following hyperparameters: 1,000 training epochs; batch size selected from \{64, 512, 4096\}; learning rate from \{0.001, 0.01\}; and dropout rate from \{0.2, 0.5\}. For the first application, we preprocess traffic sequences by limiting each to a maximum of 50 packets, with each packet truncated to 150 payload bytes and 40 header bytes, following~\cite{zhang2023tfe}. For the second application, we adopt the preprocessing and configuration from~\cite{qian2024netbench}. For the third application, we manually collect 5,000 traceroutes spanning 10 different submarine cables and split the data into training/validation/test sets using a 70\%/15\%/15\% ratio. Once the Traffic Classifier is trained, its predictions are used to derive explanations via mutual information optimization at both the byte and byte-byte levels. Explanations are represented as score masks, either vectors in $\mathbb{R}^{257}$ (for individual bytes) or matrices in $\mathbb{R}^{257 \times 257}$ (for byte-byte interactions). We rank these scores to select the Top-K bytes or interactions and evaluate explanation quality using fidelity and counterfactual accuracy. Further implementation details are provided in Appendix~\ref{app-implement}.

\begin{table*}[t!]
\small
\setlength{\tabcolsep}{1.5mm}
  \begin{tabular}{l|ll|cccc|cccc|cccc|cccc}
  \toprule
    \multirow{2}{*}{\textbf{Object}} & \multicolumn{2}{c|}{\multirow{2}{*}{\textbf{Explainer}}} & \multicolumn{4}{c|}{\textbf{ISCX-VPN}} & \multicolumn{4}{c|}{\textbf{ISCX-nonVPN}} & \multicolumn{4}{c|}{\textbf{ISCX-Tor}} & \multicolumn{4}{c}{\textbf{ISCX-nonTor}} \\
     &  & & Fid & Acc & C-Fid & C-Acc & Fid & Acc & C-Fid & C-Acc & Fid & Acc & C-Fid & C-Acc & Fid & Acc & C-Fid & C-Acc \\
     \midrule
    \multirow{15}{*}{\rotatebox{90}{\textbf{Byte Level}}} & \multirow{3}{*}{\textbf{Random}} & 1\% & 31.9 & 31.2 & 1.27 & 4.46 & 22.5 & 22.8 & 1.5 & 10.6 & 13.2 & 9.2 & 0.57 & 17.8 & 43.6 & 43.9 & 0.29 & 2.88 \\
     &  & 5\% & 35.7 & 35.7 & 4.50 & 8.30 & 25.6 & 25.8 & 4.60 & 13.4 & 14.4 & 12.6 & 3.50 & 21.3 & 42.4 & 42.7 & 1.80 & 3.40 \\
     &  & 10\% & 39.5 & 40.8 & 9.60 & 12.1 & 28.6 & 28.9 & 7.90 & 12.9 & 16.1 & 12.6 & 5.80 & 23.0 & 42.9 & 43.2 & 4.50 & 6.10 \\
     \cline{2-19}
     & \multirow{3}{*}{\textbf{\makecell[l]{Saliency\\Map}}} & 1\% & 33.8 & 35.0 & 6.40 & 9.60 & 23.3 & 23.5 & 0.76 & 10.1 & 31.6 & 22.4 & 29.3 & 42.5 & 37.9 & 38.1 & 3.00 & 4.90\\
     &  & 5\% & 69.4 & 72.0 & 50.3 & 52.2 & 40.0 & 39.5 & 14.9 & 18.7 & 32.8 & 23.0 & 28.7 & 36.2 & 31.8 & 31.8 & 12.5 & 13.2 \\
     &  & 10\% & 71.3 & 72.0 & 53.5 & 52.9 & 44.6 & 43.3 & 22.0 & 24.1 & 46.0 & 36.8 & 43.7 & 52.9 & 51.8 & 51.5 & 21.7 & 21.8 \\

    \cline{2-19}
    & \multirow{3}{*}{\textbf{SHAP}} 
      & 1\% & 33.8 & 33.8 & 0.00 & 5.73 & 22.8 & 23.0 & 0.00 & 10.4 & 9.5 & 0.00 & 0.00 & 9.5 & 44.2 & 44.5 & 0.00 & 2.63 \\
    & & 5\% & 33.8 & 33.8 & 0.00 & 5.73 & 22.8 & 23.0 & 0.00 & 10.4 & 9.5 & 0.00 & 0.00 & 9.5 & 44.2 & 44.5 & 0.00 & 2.63 \\
    & & 10\% & 33.8 & 33.8 & 0.00 & 5.73 & 22.8 & 23.0 & 0.00 & 10.4 & 9.5 & 0.00 & 0.00 & 9.5 & 44.2 & 44.5 & 0.00 & 2.63 \\
    \cline{2-19}
     & \multirow{3}{*}{\textbf{LIME}} 
      & 1\%  & \underline{80.9} & \underline{79.6} & \underline{54.1} & \underline{54.8} & \underline{60.0} & \underline{57.5} & \underline{24.1} & \underline{26.1} & \textbf{56.9} & \textbf{45.4} & \underline{42.5} & \textbf{51.7} & \underline{51.4} & \underline{51.5} & \underline{10.1} & \underline{10.8} \\
    & & 5\% & \underline{94.9} & \underline{92.4} & \underline{69.4} & \underline{70.1} & \underline{75.7} & \underline{71.1} & \underline{42.8} & \underline{40.2} & \underline{68.4} & \underline{54.6} & \underline{68.4} & \underline{73.6} & \underline{69.0} & \underline{68.8} & \underline{21.0} & \underline{20.7} \\
    & & 10\% & \underline{96.8} & \underline{93.0} & \underline{78.3} & \underline{76.4} & \underline{86.3} & \underline{79.5} & \underline{60.3} & \underline{55.7} & \underline{79.9} & \underline{66.1} & \underline{77.6} & \underline{79.9} & \underline{81.0} & \underline{80.6} & \underline{31.1} & \underline{30.4} \\

     \cline{2-19}
     & \multirow{3}{*}{\textbf{\makecell[l]{Traffic-\\Explainer}}} & 1\% & \textbf{82.8} & \textbf{81.5} & \textbf{58.6} & \textbf{56.7} & \textbf{65.3} & \textbf{65.1} & \textbf{29.1} & \textbf{33.2} & \underline{44.3} & \underline{39.1} & \textbf{49.4} & \underline{50.6} & \textbf{74.5} & \textbf{74.8} & \textbf{20.2} & \textbf{20.9} \\
     &  & 5\% & \textbf{97.5} & \textbf{92.4} & \textbf{82.8} & \textbf{79.6} & \textbf{92.4} & \textbf{84.6} & \textbf{78.5} & \textbf{74.9} & \textbf{92.0} & \textbf{77.6} & \textbf{86.8} & \textbf{88.5} & \textbf{96.1} & \textbf{95.4} & \textbf{71.1} & \textbf{70.5} \\
     &  & 10\% & \textbf{98.7} & \textbf{93.0} & \textbf{84.1} & \textbf{80.9} & \textbf{96.2} & \textbf{87.3} & \textbf{79.8} & \textbf{76.5} & \textbf{97.7} & \textbf{81.0} & \textbf{86.8} & \textbf{90.2} & \textbf{97.9} & \textbf{96.0} & \textbf{77.2} & \textbf{76.8} \\
\midrule
\midrule
    \multirow{9}{*}{\rotatebox{90}{\textbf{Byte-Byte Level}}} & \multirow{3}{*}{\textbf{Random}} & 1\% & 28.7 & 27.4 & 0.00 & 5.70 & 34.9 & 35.7 & 0.51 & 10.4 & 16.7 & 16.7 & 0.57 & \underline{18.4} & 17.7 & 17.4 & 0.83 & 2.70 \\
     &  & 5\% & 35.0 & 33.8 & 0.64 & 5.10 & 40.5 & 40.3 & 1.30 & 10.1 & 27.6 & 31.6 & 3.50 & 19.5 & 26.0 & 25.8 & 0.29 & 2.80 \\
     &  & 10\% & 38.9 & 36.9 & 1.90 & 5.10 & 52.2 & 51.9 & 2.00 & 10.4 & 31.6 & 36.2 & 4.00 & 20.7 & \underline{55.3} & \underline{55.0} & 0.42 & 2.70 \\
     \cline{2-19}
     & \multirow{3}{*}{\textbf{\makecell[l]{Self-\\Attention}}} & 1\% & \underline{93.6} & \textbf{93.6} & \underline{8.90} & \underline{10.2} & \underline{93.9} & \textbf{89.1} & \underline{8.90} & \underline{12.7} & \underline{53.5} & \underline{56.9} & \underline{9.20}  & \underline{14.4} & \underline{37.9} & \underline{38.1} & \underline{3.00} &  \underline{4.90}\\
     &  & 5\% & \underline{93.6} & \textbf{93.0} & \textbf{74.5} & \textbf{76.4} & \underline{97.0} & \textbf{89.1} & \textbf{43.5} & \textbf{44.1} & \underline{69.5} & \underline{61.5} & \underline{46.0} & \underline{52.3} & \underline{31.8} & \underline{31.8} & \underline{12.5} & \underline{13.2} \\
     &  & 10\% & \textbf{95.5} & \textbf{93.6} & \textbf{79.0} & \textbf{80.9} & \underline{98.2} & \textbf{89.6} & \textbf{56.5} & \textbf{56.7} & \underline{71.3} & \underline{61.5} & \underline{59.8} & \underline{64.4} & 51.8 & 51.5 & \underline{21.7} & \underline{21.8} \\
     \cline{2-19}
     & \multirow{3}{*}{\textbf{\makecell[l]{Traffic-\\Explainer}}} & 1\% & \textbf{97.5} & \underline{92.4} & \textbf{56.1} & \textbf{52.2} & \textbf{96.0} & \underline{87.1} & \textbf{38.7} & \textbf{34.4} & \textbf{74.7} & \textbf{63.8} & \textbf{73.0} & \textbf{69.0} & \textbf{96.8} & \textbf{94.6} & \textbf{48.5} & \textbf{47.6} \\
     &  & 5\% & \textbf{98.1} & \underline{92.4} & \underline{58.6} & \underline{58.0} & \textbf{99.0} & \underline{88.6} & \underline{37.0} & \underline{36.2} & \textbf{93.7} & \textbf{75.3} & \textbf{74.7} & \textbf{70.1} & \textbf{97.1} & \textbf{94.6} & \textbf{60.4} & \textbf{60.0} \\
     &  & 10\% & \underline{94.9} & \underline{89.2} & \underline{56.7} & \underline{58.0} & \textbf{98.7} & \underline{88.4} & \underline{38.5} & \underline{39.8} & \textbf{98.3} & \textbf{79.9} & \textbf{74.7} & \textbf{73.0} & \textbf{96.9} & \textbf{94.4} & \textbf{61.3} & \textbf{61.1}\\
     \bottomrule
    \end{tabular}
\caption{Comparison of local-instance explanation at Byte and Byte-Byte levels. The best/runner-up results under the same byte budget are in \textbf{bold} and \underline{underlined}. Our proposed Traffic-Explainer generates explanations of higher quality.}
\label{tab-local-instance-res}
\vspace{-3ex}
\end{table*}

\vspace{-2ex}
\subsection{Application Classification}
\subsubsection{Local-Instance Explanation}
In Table~\ref{tab-local-instance-res}, we evaluate Traffic-Explainer against Random/Saliency Map/SHAP/LIME for byte-level explanations, and against Random/Self-Attention for byte-byte interactional level explanations. This choice reflects the nature of each baseline: Self-Attention captures pairwise interactions and is unsuitable for byte-level explanation, while Saliency Map, SHAP, and LIME are not designed to model pairwise dependencies, making them infeasible for byte-byte interactional level explanation.

For byte-level explanations, the performance margin is significant. Traffic-Explainer requires only 5\% of bytes (approximately $256 \times 5\% \approx 10$ bytes) to explain more than half of the networking application predictions. This finding suggests that certain bytes are particularly salient and closely associated with each network class, which aligns with the principles of sparsity in deep representation learning and supports our design choice of incorporating mask regularization. Among the four baselines, the Saliency Map and LIME outperform the Random and SHAP. The Saliency Map is inferior to Traffic-Explainer because it only considers the individual importance of bytes, overlooking their cooperative effect in contributing to the final prediction. In contrast, Traffic-Explainer jointly optimizes the mask over the whole sequence to identify the Top-K important bytes, thereby inherently capturing their interactions. The weaker explanation performance of LIME relative to Traffic-Explainer is likely due to its reliance on a simple surrogate model, which struggles to approximate the complex decision boundaries of the sequence traffic classifier. In comparison, Traffic-Explainer optimizes a learned mask on the original trained model, preserving the architecture function. Moreover, as the explanation size increases (i.e., the amount of masked bytes decreases), the explanation performance increases. We also verify the efficiency of Traffic-Explainer by empirically measuring the time for explaining each individual instance. On average, Traffic-Explainer takes 2.52s/2.12s/1.90s/2.10s for generating the explanation (i.e., the byte mask) for each traffic instance across the above four datasets. Traffic-Explainer applies a mask on the input data, without imposing additional complexity, as the underlying traffic classifier remains the same.

For byte-byte level explanations, Traffic-Explainer achieves high explanation in most cases. However, it performs slightly below the Self-Attention Baseline on the ISCX-VPN and ISCX-NonVPN datasets when evaluated using Acc and C-Fid metrics. We hypothesize that this is because the Self-Attention Baseline selects Top-K byte-byte interactions from all layers of the transformer, thereby exerting a stronger influence on the model’s decision process. In contrast, Traffic-Explainer defines fixed pairwise interactions whose importance remains constant across all layers. This design choice, despite being less effective in manipulating model predictions after applying the attention mask, aligns more closely with the intrinsic characteristics of network traffic, where certain byte pairs, such as protocol indicators and IP header fields, consistently co-occur across flows within the same application or country. 

\subsubsection{Global-Class Explanation}
While local-instance explanation is valuable for understanding the model's decision-making for each network traffic, certain applications demand understanding of a group of instances by Eq.~\eqref{eq-class-explain}. In these cases, identifying common patterns that influence a group can reveal rules or insights that go beyond individual cases. Therefore, we extend our analysis to include global-class level explanations shown in Table~\ref{tab-global-class-res}. In general, our Traffic-Explainer still achieves the best explanation performance. Compared with the instance-level explanation in Table~\ref{tab-local-instance-res}, the overall decreasing quality of the class-level explanation as compared to the instance-level explanation in Table~\ref{tab-global-class-res} might be due to the increased variability inherent in aggregating explanations across a broader set of network traffic sequences.

\begin{table*}[t!]
\small
\setlength{\tabcolsep}{1.5mm}
\begin{tabular}{l|ll|cccc|cccc|cccc|cccc}
\toprule
\multirow{2}{*}{\textbf{Object}} & \multicolumn{2}{c|}{\multirow{2}{*}{\textbf{Explainer}}} & \multicolumn{4}{c|}{\textbf{ISCX-VPN}} & \multicolumn{4}{c|}{\textbf{ISCX-nonVPN}} & \multicolumn{4}{c|}{\textbf{ISCX-Tor}} & \multicolumn{4}{c}{\textbf{ISCX-nonTor}} \\
 & &  & Fid & Acc & C-Fid & C-Acc & Fid & Acc & C-Fid & C-Acc & Fid & Acc & C-Fid & C-Acc & Fid & Acc & C-Fid & C-Acc \\
 \midrule
\multirow{6}{*}{\rotatebox{90}{\textbf{Byte Level}}}
 & \multirow{3}{*}{\textbf{\makecell[l]{Saliency\\Map}}} & 1\% & \underline{35.7} & \underline{35.7} & \underline{14.0} & \underline{15.9} & \textbf{25.9} & \textbf{19.5} & \textbf{27.6} & \textbf{40.2} & \textbf{27.9} & \textbf{26.3} & \textbf{8.10} & \textbf{13.7} & \underline{24.0} & \underline{24.3} & \underline{3.46} & \underline{4.71} \\
 &  & 5\% & \underline{66.9} & \underline{69.4} & \underline{51.6} & \underline{54.1} & \underline{30.5} & \textbf{24.1} & \textbf{36.2} & \textbf{42.5} & \underline{41.5} & \underline{41.5} & \textbf{23.3} & \textbf{26.8} & \underline{58.5} & \underline{58.8} & \underline{16.5} & \underline{17.2} \\
 &  & 10\% & \underline{75.2} & \underline{75.2} & \underline{56.1} & \underline{57.3} & \underline{51.2} & \underline{48.9} & \underline{53.5} & \underline{57.5} & \underline{43.0} & \underline{41.3} & \underline{33.4} & \underline{34.9} & \underline{83.4} & \underline{84.0} & \underline{29.0} & \underline{29.3} \\
 \cline{2-19}
 & \multirow{3}{*}{\textbf{\makecell[l]{Net-\\Explainer}}} & 1\% & \textbf{76.4} & \textbf{76.4} & \textbf{40.8} & \textbf{41.4} & \underline{16.7} & \underline{12.6} & \underline{11.5} & \underline{21.8} & \underline{25.3} & \underline{24.3} & \underline{4.56} & \underline{11.1} & \textbf{45.6} & \textbf{46.1} & \textbf{3.92} & \textbf{5.04} \\
 &  & 5\% & \textbf{94.9} & \textbf{93.0} & \textbf{68.2} & \textbf{68.2} & \textbf{32.2} & \textbf{24.1} & \underline{33.9} & \underline{40.8} & \textbf{51.7} & \textbf{47.1} & \textbf{23.3} & \underline{26.3} & \textbf{81.2} & \textbf{81.6} & \textbf{29.3} & \textbf{29.3} \\
 &  & 10\% & \textbf{98.7} & \textbf{94.9} & \textbf{75.2} & \textbf{75.2} & \textbf{64.9} & \textbf{55.8} & \textbf{74.7} & \textbf{78.2} & \textbf{83.0} & \textbf{78.0} & \textbf{47.3} & \textbf{47.1} & \textbf{91.5} & \textbf{91.2} & \textbf{52.6} & \textbf{52.6} \\
 \bottomrule
\end{tabular}
\caption{Comparison of global class explanation. Traffic-Explainer exhibits higher quality in its generated explanation. We omit results for Random due to its inferior performance in Table~\ref{tab-local-instance-res}, and LIME/SHAP due to their focus on local explanations.}
\label{tab-global-class-res}
\vspace{-6ex}
\end{table*}

\begin{figure*}[t!]
    \centering
    \includegraphics[width=1.0\textwidth]{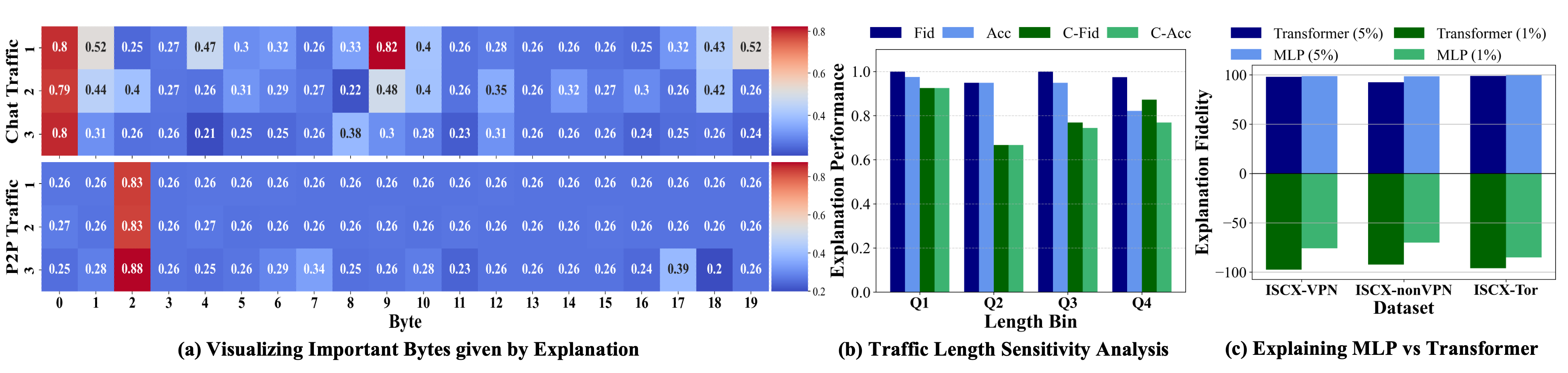}
    \vspace{-6ex}
    \caption{\textbf{(a)} By visualizing the byte importance for each networking traffic sequence in Chat and P2P applications, network operators can assess how the identified critical bytes correspond to the actual content carried by the traffic, thereby validating the predicted application.
    \textbf{(b)} We observe a slightly decreasing traffic explanation quality as the sequence length grows.
    \textbf{(c)} Traffic-Explainer achieves consistently higher explanation performance across both MLP/Transformer-based network classifiers.}
    \label{fig-analysis}
\end{figure*}

\subsubsection{Explanation Visualization}
After quantitatively validating the quality of explanations generated by our proposed Traffic-Explainer, we qualitatively visualize the byte importance scores for three network sequences from Chat and P2P applications. Figure~\ref{fig-analysis}(a) illustrates a distinct byte importance pattern, where the $0^{\text{th}}/1^{\text{st}}/9^{\text{th}}$ bytes are crucial for identifying traffic of application Chat, while the $2^{\text{nd}}$ byte plays the most significant role in distinguishing traffic of P2P application. The explanation also differs across different flow sequences, even though they belong to the same traffic application. 

\subsubsection{Sensitivity Analysis of Traffic Length}
We conduct a sensitivity analysis to evaluate how the length of network traffic sequences impacts explanation quality. Specifically, on the ISCX-VPN dataset, we group traffic sequences into four bins by the total number of bytes (combining both header and payload) for each network traffic sequence and measure the average explanation performance in Figure~\ref{fig-analysis}(b). We observe a slight decrease in explanation quality as sequence length increases, although the pattern is not strictly monotonic. We attribute this to the increased difficulty of identifying necessary bytes within longer sequences. As the sequence grows, the larger number of interacting byte pairs makes it harder for Traffic-Explainer to recognize the contribution of specific bytes.

\subsubsection{Model-Agnostic Nature of Traffic-Explainer on MLP/Transformer Classifiers}
To demonstrate the generality of our Traffic-Explainer, we further apply it to the MLP-based traffic classifier and compare its explanation fidelity against the Transformer-based counterpart under 1\%/5\% byte budgets across three datasets. In Figure~\ref{fig-analysis}(c), Traffic-Explainer achieves similarly high explanation performance between MLP and transformer backbone, confirming its model-agnostic effectiveness.

\subsection{Traffic Localization}

In this section, we present the second application of Traffic-Explainer: traffic localization. This use case highlights both the interpretability benefits and potential adversarial risks of applying our Traffic-Explainer. When users access the internet from specific geographic regions, such as a particular country, their traffic patterns inherently reflect regional infrastructure and locality characteristics~\cite{malecki2002economic}. Traffic-Explainer captures these patterns by automatically identifying the most influential bytes within a traffic sequence that contribute to predicting the country of origin. Following the experimental setting of network application classification in Table~\ref{tab-local-instance-res}, we present the explanation performance of traffic localization in Table~\ref{tab-country-perform} in Appendix~\ref{app-result}. Overall, our Traffic-Explainer continues to demonstrate superior explanation performance.

\begin{figure}[t!]
    \centering
    \includegraphics[width=0.5\textwidth]{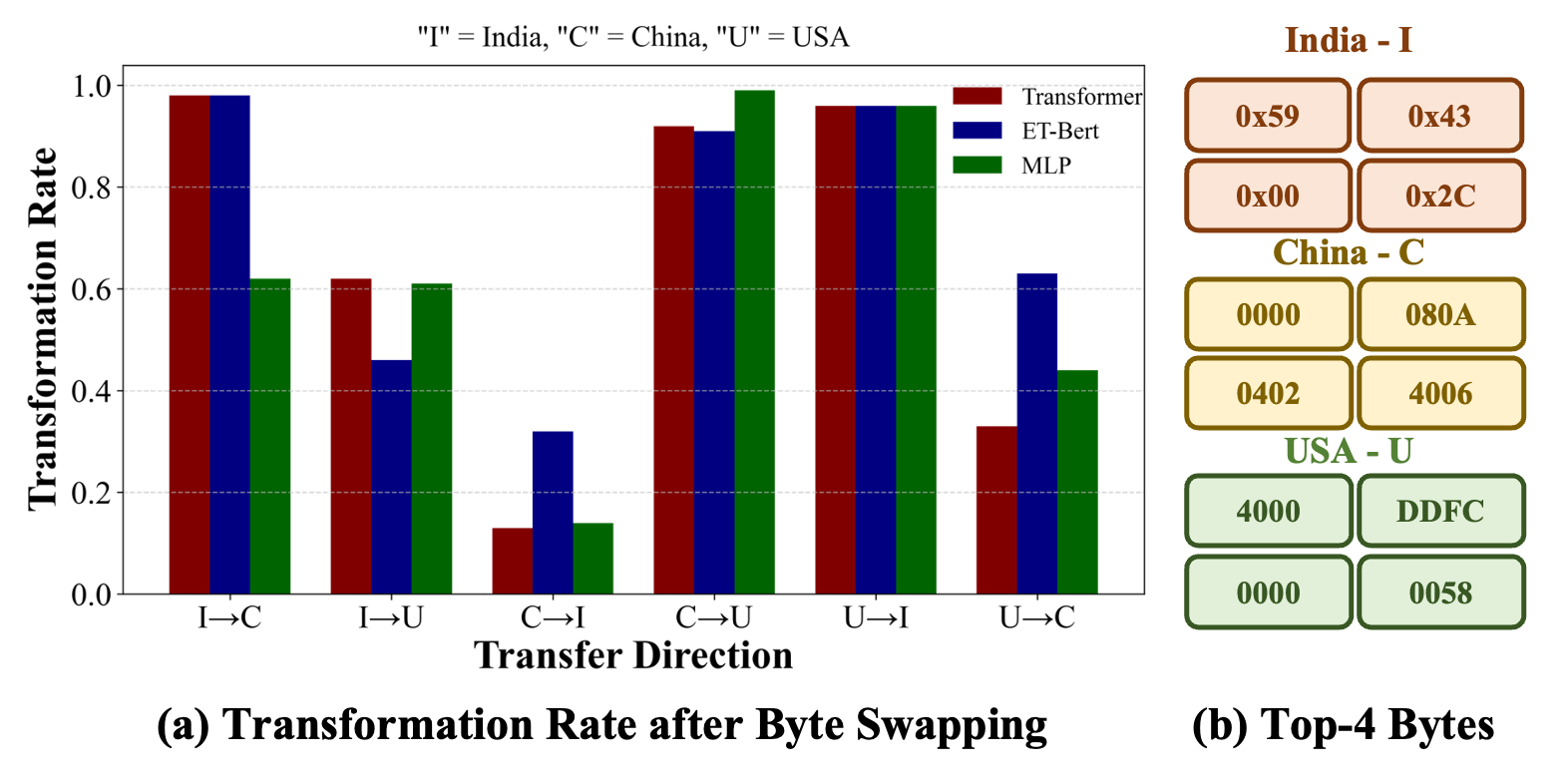}
    \vspace{-5ex}
    \caption{(a) After swapping Top 10\% of explanatory bytes by Traffic-Explainer between traffic sequences from different country locations, all classifiers (Transformer/ET-BERT~\cite{lin2022bert}/MLP) exhibit strong class transformation rates. The fact that the transformation also works on models other than the one used for explanation suggests that the identified bytes are truly causal for predicting the country, not just artifacts of the Transformer model. (b) Visualizing Top-4 most important bytes for each country-level traffic class.}
    \label{fig-transfer}
    \vspace{-2ex}
\end{figure}

At the global class level, we aim to evaluate the causal and transferable nature of the identified explanations. We perform a byte-swapping experiment across traffic sequences occurring at different country locations. Specifically, we swap the top 10\% most important bytes, identified by Traffic-Explainer, between traffic flows belonging to different countries and examine how this affects the output of various classifiers, including Transformer, ET-Bert~\cite{lin2022bert}, and MLP. As shown in Figure~\ref{fig-transfer}(a), this manipulation consistently causes the classifiers to change their predictions toward the target country associated with the swapped bytes, demonstrating a strong transformation rate. 
\textit{The success of these transformations, even when using classifiers different from the one used for applying Traffic-Explainer during explanation generation, indicates that the identified bytes are not merely model-specific explanations but causal features that truly characterize country-level traffic patterns.} Note that our byte-swapping operation involves exchanging 16-bit hex values between two valid sequences, so the checksums remain correct, and the network traffic is still valid in the real world after swapping bytes. Figure~\ref{fig-analysis}(b) visualizes the most influential bytes for representative countries such as India, China, and the USA. These visualizations offer interpretable insights into regional traffic signatures and can inform more precise, geo-aware networking services. Practical applications include region-specific content delivery, targeted advertising, geolocation-based authentication, geofencing, and optimized routing in multiplayer gaming environments~\cite{rusek2020routenet}. Beyond technical utility, Traffic-Explainer also provides value from a social-good perspective. In regions with government surveillance or censorship, this tool helps users protect their privacy by identifying and obfuscating location-revealing byte patterns. For instance, journalists in sensitive areas could use Traffic-Explainer to deduce information about vulnerable bytes and prevent their actual location from being detected, reducing surveillance risks and cyber-attacks.

%

\subsection{Network Cartography}
In this section, we present the third application of Traffic-Explainer: network cartography. Specifically, we aim to validate whether the traceroute-to-submarine cable mapping correctly identifies the key hop in the round-trip time sequence for accurate mapping. As network signals travel through physical cables, the RTT is measured at each hop. Longer physical cables generally result in higher RTTs at the corresponding hops, and submarine cables, which are significantly longer than terrestrial ones, exhibit distinct RTT patterns. This observation leads us to frame the traceroute mapping task as a submarine cable classification problem, where each traceroute sequence is associated with the specific submarine cable it traverses. To enhance transparency, we apply Traffic-Explainer to identify the RTT hops most determinative in the classification decision. These identified hops are cross-validated against the ground-truth RTT sequence: if they align with the hops exhibiting significant RTT spikes, it confirms the reliability of our mapping model.

\begin{figure}[t!]
    \centering
    \includegraphics[width=0.5\textwidth]{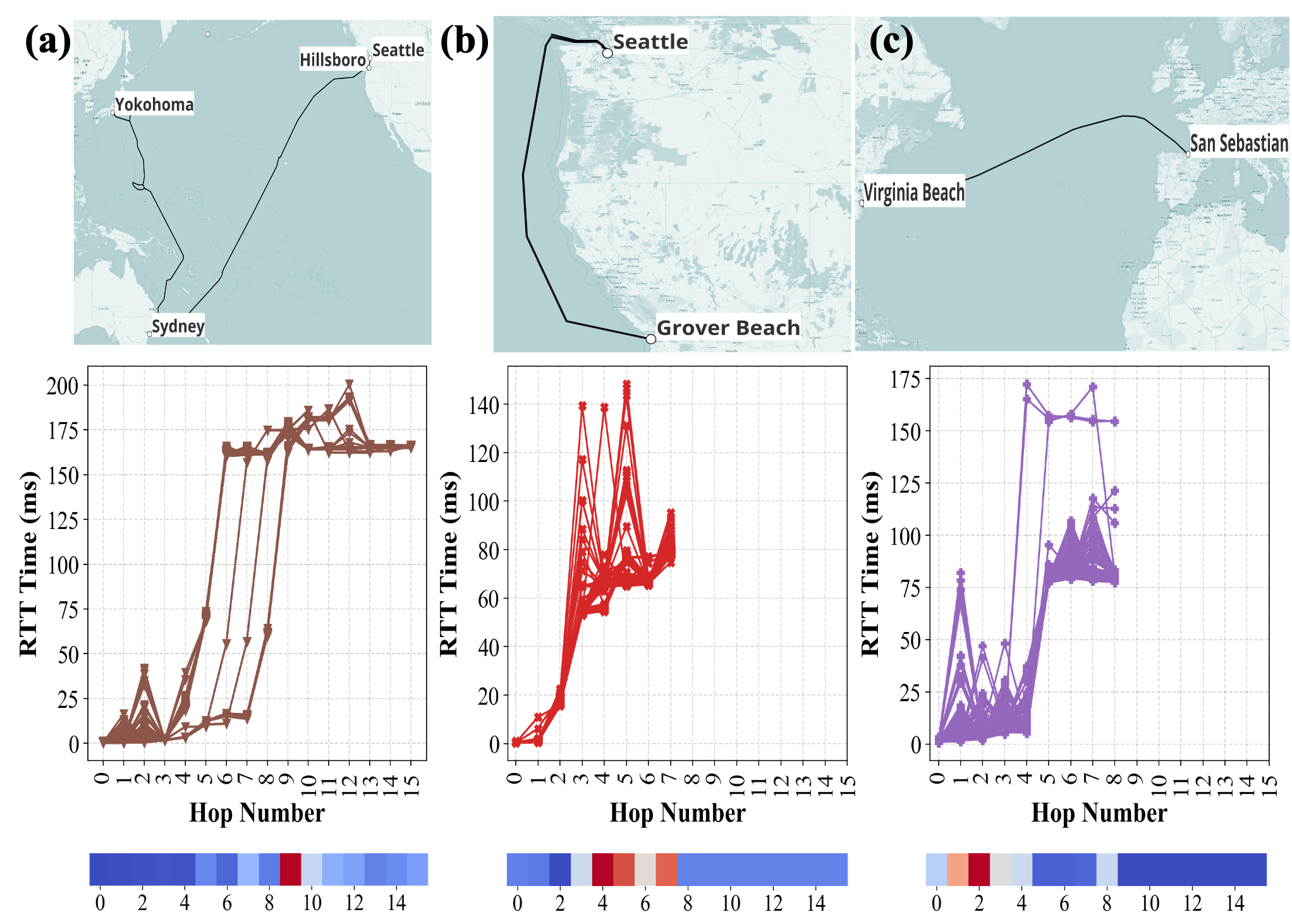}
    \vspace{-5ex}
    \caption{Top - Three distinct submarine cables used for transmitting three groups of traffic traceroute: (a) Seattle, US $\rightarrow$ Yokohama, Japan; (b) Seattle, US $\rightarrow$ San Jose, US; (c) Virginia Beach, US $\rightarrow$ San Sebastian, France. Middle - Sequences of RTT measurements for different traceroute paths. Bottom - Explanation by identifying the most responsible cable for inferring the physical route of a traffic path.}
    \label{fig-rtt}
    \vspace{-4ex}
\end{figure}

We collaborate with domain experts from Link Oregon to collect 5,000 traceroute measurements traversing 15 distinct submarine cables using tools such as RIPE Atlas and CAIDA Ark~\cite{gharaibeh2017look}. Each traceroute consists of a round-trip time sequence that reflects the physical path of the probe, including both terrestrial and submarine cable segments. Figure~\ref{fig-rtt} highlights three representative traceroutes, with the corresponding RTT sequences shown below each. Notably, the RTT patterns exhibit distinct sequential structures across different submarine cables. Leveraging these patterns, our transformer-based traffic classifier demonstrates exceptional classification performance, as shown in Table~\ref{tab:class_accuracy_transposed}. To ensure the network cartography (mapping traceroute to physical submarine cable) is based on the distinct RTT spike patterns for each traceroute, we apply Traffic-Explainer to automatically identify the most influential RTT hop responsible for the traceroute mapping. We visualize the heatmap scores for each traceroute hop under the RTT sequence in Figure~\ref{fig-rtt}. The darker red regions, indicating higher importance, align well with the RTT spikes. This alignment confirms that our mapping approach successfully captures RTT spikes as discriminative features for characterizing the underlying physical path.


%% file: Conclusion.tex
\section{Conclusion}
Despite the successful adoption of deep learning-powered models in network traffic classification, most of them focus on boosting performance without considering the underlying reasons behind their decision-making process, which jeopardizes the transparent adoption of DL solutions by network operators. This motivates us to introduce an explanation framework, Traffic-Explainer, to explain traffic predictions by identifying the important units and unit-unit interactions in given sequences. Extensive real-world evaluations validate the effectiveness of Traffic-Explainer in both local-instance and global-class predictions, demonstrating its potential to advance transparent network management.


%% file: Appendix.tex
\begin{table*}[t!]
\begin{tabular}{llccccc}
\toprule
\textbf{Dataset} & \textbf{Type} & \# \textbf{Train/Val/Test Seq.} & \# \textbf{Packet} & \# \textbf{Byte} & \textbf{Task} & \# \textbf{Label} \\
\midrule
\multirow{2}{*}{\textbf{\makecell[l]{ISCX-\\VPN}}} & Header & \multirow{2}{*}{1,231/154/157} & 34.20$\pm$15.23 & 1,351$\pm$612.4 Byte & \multirow{2}{*}{Application Classification} & \multirow{2}{*}{6} \\
 & Payload &  & 24.62$\pm$18.99 & 2,828$\pm$2,742 Byte &  &  \\
 \midrule
\multirow{2}{*}{\textbf{\makecell[l]{ISCX-\\NonVPN}}} & Header & \multirow{2}{*}{3,140/392/395} & 25.23$\pm$15.66 & 1,009$\pm$626.3 Byte & \multirow{2}{*}{Application Classification} & \multirow{2}{*}{6} \\
 & Payload &  & 14.88$\pm$16.05 & 1,641$\pm$2,057 Byte &  &  \\
 \midrule
\multirow{2}{*}{\textbf{\makecell[l]{ISCX-\\Tor}}} & Header & \multirow{2}{*}{1,354/169/174} & 43.03$\pm$14.92 & 1,721$\pm$596.6 Byte & \multirow{2}{*}{Application Classification} & \multirow{2}{*}{8} \\
 & Payload &  & 40.94$\pm$17.02 & 6,093$\pm$2,543 Byte &  &  \\
 \midrule
\multirow{2}{*}{\textbf{\makecell[l]{ISCX-\\NonTor}}} & Header & \multirow{2}{*}{19,179/2,398/2,400} & 26.00$\pm$16.77 & 1,040$\pm$670.7 Byte & \multirow{2}{*}{Application Classification} & \multirow{2}{*}{8} \\
 & Payload &  & 13.60$\pm$17.06 & 1,740$\pm$2,411 Byte &  &  \\
 \midrule
\multirow{2}{*}{\textbf{IOS}} & Header & \multirow{2}{*}{776,156/97,803/97,803} & / & 25.74$\pm$2.740 Byte & \multirow{2}{*}{Traffic Localization} & \multirow{2}{*}{3} \\
 & Payload &  & / & 28.27$\pm$29.93 Byte &  &  \\
 \midrule
\multirow{2}{*}{\textbf{Android}} & Header & \multirow{2}{*}{1,086,909/135,691/135,692} & / & 25.66$\pm$2.670 Byte & \multirow{2}{*}{Traffic Localization} & \multirow{2}{*}{3} \\
 & Payload &  & / & 28.84$\pm$31.04 Byte &  & \\
  \midrule
\multirow{2}{*}{\textbf{Traceroute}} & \multirow{2}{*}{RTT} & \multirow{2}{*}{10\%/10\%/80\%} & / & / & \multirow{2}{*}{Network Cartography} & \multirow{2}{*}{15} \\
 &  &  & / & / &  & \\
 \bottomrule
\end{tabular}
\centering
\caption{Statistics of Network Traffic Sequence Datasets. The first four for application classification are from~\cite{zhang2023tfe}, while the next two for traffic localization are collected from~\cite{qian2024netbench}. The last one for traceroute is collected using RIPE Atlas probes.}
\label{tab-dataset}
\end{table*}

\section{Datasets}\label{app-data}
Seven datasets are used to verify our proposed Traffic-Explainer.

\begin{itemize}[leftmargin=*]
    \item \textbf{Application 1 - ISCX VPN/nonVPN}~\cite{zhang2023tfe}: A public traffic dataset including ISCX-VPN/non-VPN datasets. The ISCX-VPN is collected over virtual private networks (VPNs), used for accessing some blocked websites or services, and is difficult to recognize due to the obfuscation technology. Conversely, the traffic in ISCX-nonVPN is regular and not collected over VPNs.

    \item \textbf{Application 1 - ISCX Tor/NonTor}~\cite{zhang2023tfe}: ISCX Tor-nonTor is a public dataset, and the ISCX-Tor dataset is collected over the onion router, whose traffic can be difficult to trace. Besides, ISCX-nonTor is also regular and not collected over the onion router.

    \item \textbf{Application 2 - IOS/Android}~\cite{qian2024netbench}: The Cross-Platform dataset comprises user-generated data for 215 Android and 196 iOS apps. The iOS apps were gathered from the top 100 apps in the US, China, and India App Stores. The Android apps originate from the top 100 apps in the Google Play Store in the US and India, plus from the top 100 apps of the Tencent MyApps and 360 Mobile Assistant stores, as Google Play is unavailable in China. Each app was executed between three and ten minutes while receiving real user inputs. We use this dataset to evaluate our method's performance with user-generated data and the performance between different operating systems.

    \item \textbf{Application 3 - Network Traceroute Data}, we collect traceroutes from predefined sources to destinations. Using RIPE Atlas probes~\cite{holterbach@ripe}, we select target countries and direct probes to international servers. Over a three-day period, we collected 5000 unique source-to-destination traceroutes using 15 submarine cables, including Seattle, US to Yokohama, Japan; Seattle, US to San Jose, US; and Virginia Beach, US to San Sebastian, France.
\end{itemize}

We follow~\cite{zhang2023tfe} to preprocess the first four synthetic traffic flow datasets and follow~\cite{qian2024netbench} to preprocess the last two real-world traffic flow datasets. The comprehensive statistics of each dataset is presented in Table~\ref{tab-dataset}. Note that to avoid any overfitting bias, unlike~\cite{zhang2023tfe}, we also add the validation split following ratio: 80\%/10\%/10\%.

\section{Computation of Evaluation Metric}\label{app-evaluation}
As our contributions involve both a scalable traffic classifier and a trustworthy traffic explainer, our evaluation metrics should also consider these two aspects: one for evaluating the traffic classification performance and the other for evaluating the explanation quality. We use Accuracy (Acc) and F1-macro to evaluate the traffic classification performance to avoid the bias caused by the imbalance of traffic flow instances. To evaluate the explanation quality, we first apply Traffic-Explainer to select the top-K important bytes, and then we re-calculate the model predictions with the updated flow sequences by either removing or keeping those selected important bytes. For $i^{\text{th}}$ instance, assuming $\mathcal{Y}^{i, *}$ represents its ground-truth label, $\mathcal{Y}^i$ represents its predicted label with the original traffic flow sequence, $\mathcal{Y}^{F, i}$ is its updated prediction after keeping only the Top-K important bytes and masking all others, and $\mathcal{Y}^{CF, i}$ is its updated prediction after removing the Top-K important bytes and keeping all others. Then, the following four explanation evaluation metrics can be calculated as:

\begin{itemize}[leftmargin=*]
    \item \textbf{Fidelity (Fid)}: The percentage of updated predictions that equal the original model predictions after keeping only the Top-K important bytes: $\text{Fid} = \frac{1}{N} \sum_{i=1}^{N} \mathbb{I}(\mathcal{Y}^{F, i} = \mathcal{Y}^{i, *})$.
    
    \item \textbf{Accuracy (Acc)}: The percentage of updated predictions that equal the ground-truth labels after keeping only the Top-K important bytes: $\text{Acc} = \frac{1}{N} \sum_{i=1}^{N} \mathbb{I}(\mathcal{Y}^{F, i} = \mathcal{Y}^{i, *})$
    
    \item \textbf{Counterfactual-Fidelity (C-Fid)}: The percentage of updated predictions that equal the original model predictions after removing the Top-K important bytes: $\text{C-Fid} = \frac{1}{N} \sum_{i=1}^{N} \mathbb{I}(\mathcal{Y}^{CF, i} = \mathcal{Y}^{i, *})$.
    
    \item \textbf{Counterfactual-Accuracy (C-Acc)}: The percentage of updated predictions that equal the ground-truth labels after removing the Top-K important bytes: $\text{C-Acc} = \frac{1}{N} \sum_{i=1}^{N} \mathbb{I}(\mathcal{Y}^{CF, i} = \mathcal{Y}^{i, *})$
    
\end{itemize}





\begin{table*}[t!]
\begin{tabular}{llcccccccc}
\toprule
\multirow{2}{*}{\textbf{Explainer}} & \multirow{2}{*}{\textbf{Budget}} & \multicolumn{4}{c}{\textbf{IOS}} & \multicolumn{4}{c}{\textbf{Android}} \\
 &  & Fid & Acc & C-Fid & C-Acc & Fid & Acc & C-Fid & C-Acc \\
 \hline
\multirow{3}{*}{Random} & 1\% & 39.63\% & 39.46\% & 2.60\% & 3.93\% & 18.12\% & 18.05\% & 1.18\% & 1.62\% \\
 & 5\% & 40.71\% & 40.32\% & 2.88\% & 4.07\% & 24.43\% & 24.39\% & 1.81\% & 2.28\% \\
 & 10\% & 50.62\% & 50.33\% & 3.79\% & 5.04\% & 31.38\% & 31.09\% & 2.20\% & 2.66\% \\
 \hline
\multirow{3}{*}{Saliency Map} & 1\% & \textbf{72.51\%} & \textbf{71.83\%} & \textbf{15.91\%} & \textbf{16.52\%} & \underline{75.42\%} & \underline{75.34\%} & \textbf{13.43\%} & \textbf{14.09\%} \\
 & 5\% & 92.64\% & 91.68\% & 39.97\% & 40.40\% & 90.17\% & 89.87\% & 22.30\% & 22.67\% \\
 & 10\% & 95.44\% & 94.39\% & 46.64\% & 46.95\% & 94.45\% & 94.03\% & 25.75\% & 25.96\% \\
 \hline
\multirow{3}{*}{Traffic-Explainer} & 1\% & \underline{45.43\%} & \underline{44.47\%} & \underline{5.33\%} & \underline{5.70\%} & \textbf{76.85\%} & \textbf{76.13\%} & \underline{11.37\%} & \underline{11.46\%} \\
 & 5\% & \textbf{94.61\%} & \textbf{92.93\%} & \underline{39.05\%} & \underline{38.96\%} & \textbf{96.55\%} & \textbf{95.73\%} & \textbf{33.39\%} & \textbf{33.26\%} \\
 & 10\% & \textbf{97.92\%} & \textbf{96.43\%} & \textbf{62.28\%} & \textbf{62.34\%} & \textbf{99.16\%} & \textbf{98.57\%} & \textbf{44.91\%} & \textbf{44.80\%}\\
 \bottomrule
\end{tabular}
\centering
\caption{Comparison of local-instance explanation at Byte levels for traffic localization task. Our proposed Traffic-Explainer generates explanations of higher quality. Traffic-Explainer achieves better explanation than Saliency Map.}
\label{tab-country-perform}
\vspace{-4ex}
\end{table*}

\section{Implementation Details of Network Cartography}\label{app-implement}

The Traffic-Explainer is further used to address the traffic mapping challenge in network cartography, where the objective is to establish cross-layer dependencies between logical IP routes and the underlying physical infrastructure. We collaborate with domain experts to collect traceroute data using tools such as RIPE Atlas and CAIDA Ark~\cite{gharaibeh2017look}. After preprocessing, we employ IP geolocation services to associate IP addresses with precise physical locations, incorporating latitude and longitude coordinates~\cite{Hauff@qgis}. Additionally, we analyze traceroutes that traverse both terrestrial and submarine fiber-optic networks, classifying them based on RTT characteristics. The key insight is that traffic flowing through the same physical link—such as a specific submarine cable—should exhibit similar RTT patterns. Traffic-Explainer helps uncover these patterns, providing a holistic view of how different fiber-optic pathways influence traffic behavior and performance. We detail in four steps below. 

\textbf{Step 1: Data Collection. } The data collection process involves gathering traceroutes from predefined source locations to designated destinations. Using RIPE Atlas probes~\cite{holterbach@ripe}, we select target countries and direct probes to a consistent set of international servers. For example, in a case study analyzing transpacific traffic, we selected the United States as the source and Japan as the destination. Over a three-day period, we collected nearly 5000 traceroute data, including Palo Alto, US to Yokohama, Japan; Rockville, Maine, US to Amsterdam, Netherlands; and Vancouver, Canada to Washington, US. These traceroutes traverse both submarine and terrestrial fiber-optic networks, enabling cross-layer analysis of network infrastructure.

\textbf{Step 2: Data Preprocessing and Geolocation.} Once collected, the traceroute data is preprocessed to extract key hop-level details, including hop number, IP address, and source-destination pair. To assess the consistency of multiple traceroutes between the same endpoints, we visualize RTT variations across different probe times (see middle line charts in Figure~\ref{fig-rtt}). These visualizations help identify potential deviations in traffic flow and latency variations that indicate the use of different physical network paths -- especially submarine cables, as indicated by the sharp increase in RTTs.

For geolocation, we use services such as IPGeolocation, IPLocation.net, and IPinfo to determine the approximate physical location of each hop~\cite{Dan@traceroute}. The extracted geographic coordinates are then mapped using QGIS, providing a spatial representation of network paths. This geolocation data is crucial for identifying submarine and terrestrial infrastructure used by the traffic, helping establish physical dependencies in network routing.

\textbf{Step 3: Mapping Traceroutes to Submarine Infrastructure.} To map traceroutes to submarine cable infrastructure, we analyze RTT characteristics and geolocation data. First, we identify RTT spikes along traceroutes, as submarine cables typically introduce higher latency due to increased physical distance and transmission delay. 
Next, we overlay the preprocessed traceroute paths onto QGIS, mapping their geographic coordinates to known submarine cable locations. Additionally, we perform DNS reverse lookups to correlate IP addresses with submarine cable domains, further validating the inferred physical pathways. By linking domain names to submarine cable infrastructure, we identify the most probable submarine routes taken by the traffic, which are collected as a few labeled training sets for instantiating the semi-supervised submarine cable mapping/classification based on RTT characteristics. The hypothesis is that traffic flowing through the same submarine cable exhibits similar RTT patterns~\cite{Sengupta@rtt}. This classification enables a more refined understanding of how submarine cables influence network performance, congestion, and routing decisions.

\textbf{Step 4: Traffic-Explainer for Route Interpretation.} The mapped traceroute data is further analyzed using Traffic-Explainer, which enhances the interpretation of traffic flow through submarine and terrestrial networks. By overlaying traceroutes onto submarine cable maps, we verify their alignment with known physical routes. Visualizing RTT spikes along the path provides deeper insights into where traffic transitions onto submarine cables, highlighting latency variations across terrestrial vs. submarine network paths.

More broadly, by leveraging Traffic-Explainer, network operators improve network observability and gain a deeper understanding of how submarine cables impact traffic latency, routing behaviors, and  performance. This approach significantly enhances their ability to map logical-layer traffic onto the underlying physical infrastructure, advancing the field of network cartography.

\begin{table*}[t!]
\centering
\small
\begin{tabular}{|c|c c c c c c c c c c c c c c c c|}
\hline
\textbf{Metric} & \textbf{1} & \textbf{2} & \textbf{3} & \textbf{4} & \textbf{5} & \textbf{6} & \textbf{7} & \textbf{8} & \textbf{9} & \textbf{10} & \textbf{11} & \textbf{12} & \textbf{13} & \textbf{14} & \textbf{15} & \textbf{16} \\
\hline
Correct   & 1   & 202 & 110 & 226 & 233 & 211 & 235 & 103 & 428 & 94  & 226 & 107 & 237 & 402 & 394 & 469 \\
Total     & 239 & 202 & 110 & 227 & 233 & 213 & 255 & 103 & 428 & 95  & 232 & 107 & 237 & 402 & 397 & 469 \\
Accuracy (\%) & 0.42 & 100.00 & 100.00 & 99.56 & 100.00 & 99.06 & 92.16 & 100.00 & 100.00 & 98.95 & 97.41 & 100.00 & 100.00 & 100.00 & 99.24 & 100.00 \\
\hline
\end{tabular}
\caption{Network cartography performance of traceroute mapping across different source-destination pairs.}
\label{tab:class_accuracy_transposed}
\end{table*}

\section{Additional Results}\label{app-result}

\subsection{Networking Cartography Performance}
We collected approximately 5,000 traceroute measurements over a three-day period across 15 unique source-destination pairs, with each pair associated with a specific submarine cable. To train a traceroute mapping model that classifies a given traceroute to its corresponding source-destination pair, we labeled 5\% of the traceroutes within each group and conducted a classification experiment. As shown in Table~\ref{tab:class_accuracy_transposed}, the model achieves high accuracy, confirming a strong correlation between physical cable paths and logical RTT patterns. These results further motivate our RTT-based explanation learning, where we derive a mask to automatically identify the RTT hops most indicative of the underlying physical cable. In Figure~\ref{fig-rtt}, Traffic-Explainer successfully identifies the most characteristic hop, typically corresponding to the largest RTT jump, highlighting the transition point associated with the submarine cable


\section{Real-World Implications}\label{app-implication}

A natural question to consider is: {\em what are the real-world implications and operational benefits of deploying the Traffic-Explainer framework}? In this section, we describe how Traffic-Explainer provides operational value to network operators and researchers alike across different operational settings. The framework is designed to offer transparent, interpretable insights into deep learning model predictions, thereby closing a critical gap between high-performing but opaque models and the need for accountability and transparency in production network environments~\cite{lundberg2017shap,zhang2023tfe}.

From a general perspective, Traffic-Explainer enhances transparency in deep learning-powered networking systems by providing explanations at both the instance and class levels. For network operators performing monitoring, securing, and optimizing networks, such explanations are essential. In fact, it allows operators to understand not only {\em what} the model predicts but also {\em why} it makes that decision~\cite{lundberg2017shap}. This transparency supports tasks like debugging misclassifications, detecting anomalies, validating compliance with policy or legal requirements, and even auditing model behavior for drift over time. By highlighting the most influential input features (e.g., bytes in packet sequences or RTT hops in traceroute data), Traffic-Explainer acts as a diagnostic tool, empowering operators to make informed, actionable decisions~\cite{pujol2021netxplain}.

\subsection{Traffic Application Classification}
In the context of network application classification, Traffic-Explainer allows operators to identify which specific byte patterns within traffic flows are driving classification decisions. This enables deeper traffic forensics. For example, if a flow is classified as ``YouTube” operators can inspect whether the decision was influenced by protocol-specific headers, certificate fingerprints, or payload markers~\cite{zhang2023tfe}. This insight is valuable for enforcing application-specific policies, detecting evasion techniques (e.g., apps mimicking other protocols), or even discovering previously unknown traffic signatures. 
The byte-level visibility into model decisions also reduces the reliance on operator know-how (e.g., hand-engineered features), which are traditionally brittle and require constant manual updates.

\subsection{Traffic Country Localization}
For network country localization, Traffic-Explainer helps surface the regional indicators embedded in traffic data that contribute to country-level classification. 
This is especially relevant in scenarios where regulatory compliance or geopolitical constraints are in play~\cite{jiang2016ip}. For instance, operators can verify whether certain packets are correctly identified as originating from restricted or sanctioned regions, based on explainable byte patterns rather than blind model outputs. 
Beyond compliance, this capability can also be used to detect traffic masking techniques, such as VPN or proxy usage, and to uncover implicit privacy risks. 
Importantly, Traffic-Explainer can aid users in understanding which parts of their traffic expose location information, thus enabling the design of more privacy-preserving communication strategies (e.g., privacy-preserving protocols).

\subsection{Network Cartography}
In the case of network cartography, Traffic-Explainer provides interpretable mappings between logical observations (traceroute sequences) and physical infrastructure (such as submarine cables)~\cite{durairajan2015intertubes,wang2025sigcomm}. 
By identifying the RTT hops most responsible for inferring a traffic path's physical route, Traffic-Explainer enables a new level of visibility into network topology and routing behavior. 
This can assist operators in diagnosing path anomalies, assessing the resilience of routing paths, and planning for failure scenarios (both benign and malicious). 
The ability to highlight specific RTT spikes corresponding to cable segments (e.g., transoceanic hops) is particularly useful for submarine risk assessment and capacity planning.

\section{Ethical Concern and Mitigation}\label{app-ethical}
Traffic-Explainer is designed to enhance transparency in network traffic classification, not to function as a surveillance tool. However, its capability to automatically uncover fine-grained packet features and traceroute paths may inadvertently expose sensitive user information. For instance, traffic localization could reveal a user’s approximate geographic location (e.g., the country). Additionally, the discovery of country-specific byte patterns could be adversarially exploited to spoof or falsify location information. These risks are particularly concerning in sensitive regions or under strict regulatory regimes, where misuse could lead to unwarranted surveillance or the de-anonymization of individuals. To mitigate these concerns, we propose several safeguards: (1) User consent mechanisms to ensure detailed traffic analysis is performed only with explicit permission; (2) Limiting geographic inference granularity, providing only coarse region-level hints rather than precise locations; (3) Masking or omitting identifiable byte patterns from any publicly released outputs; and (4) Institutional oversight, such as ethics board review or compliance audits, for any system deployment. By implementing these safeguards and adhering to strict ethical standards, researchers and network operators can responsibly leverage Traffic-Explainer’s transparency benefits without compromising user privacy and informing risks.

%% file: main.bbl

\begin{thebibliography}{49}


\ifx \showCODEN    \undefined \def \showCODEN     #1{\unskip}     \fi
\ifx \showDOI      \undefined \def \showDOI       #1{#1}\fi
\ifx \showISBNx    \undefined \def \showISBNx     #1{\unskip}     \fi
\ifx \showISBNxiii \undefined \def \showISBNxiii  #1{\unskip}     \fi
\ifx \showISSN     \undefined \def \showISSN      #1{\unskip}     \fi
\ifx \showLCCN     \undefined \def \showLCCN      #1{\unskip}     \fi
\ifx \shownote     \undefined \def \shownote      #1{#1}          \fi
\ifx \showarticletitle \undefined \def \showarticletitle #1{#1}   \fi
\ifx \showURL      \undefined \def \showURL       {\relax}        \fi
\providecommand\bibfield[2]{#2}
\providecommand\bibinfo[2]{#2}
\providecommand\natexlab[1]{#1}
\providecommand\showeprint[2][]{arXiv:#2}

\bibitem[Abdullahi et~al\mbox{.}(2022)]%
        {abdullahi2022detecting}
\bibfield{author}{\bibinfo{person}{Mujaheed Abdullahi}, \bibinfo{person}{Yahia Baashar}, \bibinfo{person}{Hitham Alhussian}, \bibinfo{person}{Ayed Alwadain}, \bibinfo{person}{Norshakirah Aziz}, \bibinfo{person}{Luiz~Fernando Capretz}, {and} \bibinfo{person}{Said~Jadid Abdulkadir}.} \bibinfo{year}{2022}\natexlab{}.
\newblock \showarticletitle{Detecting cybersecurity attacks in internet of things using artificial intelligence methods: A systematic literature review}.
\newblock \bibinfo{journal}{\emph{Electronics}} \bibinfo{volume}{11}, \bibinfo{number}{2} (\bibinfo{year}{2022}), \bibinfo{pages}{198}.
\newblock


\bibitem[Adadi and Berrada(2018)]%
        {adadi2018peeking}
\bibfield{author}{\bibinfo{person}{Amina Adadi} {and} \bibinfo{person}{Mohammed Berrada}.} \bibinfo{year}{2018}\natexlab{}.
\newblock \showarticletitle{Peeking inside the black-box: a survey on explainable artificial intelligence (XAI)}.
\newblock \bibinfo{journal}{\emph{IEEE access}}  \bibinfo{volume}{6} (\bibinfo{year}{2018}).
\newblock


\bibitem[Alangari et~al\mbox{.}(2023)]%
        {alangari2023exploring}
\bibfield{author}{\bibinfo{person}{Nourah Alangari}, \bibinfo{person}{Mohamed El~Bachir~Menai}, \bibinfo{person}{Hassan Mathkour}, {and} \bibinfo{person}{Ibrahim Almosallam}.} \bibinfo{year}{2023}\natexlab{}.
\newblock \showarticletitle{Exploring evaluation methods for interpretable machine learning: A survey}.
\newblock \bibinfo{journal}{\emph{Information}} \bibinfo{volume}{14}, \bibinfo{number}{8} (\bibinfo{year}{2023}), \bibinfo{pages}{469}.
\newblock


\bibitem[Amann et~al\mbox{.}(2020)]%
        {amann2020explainability}
\bibfield{author}{\bibinfo{person}{Julia Amann}, \bibinfo{person}{Alessandro Blasimme}, \bibinfo{person}{Effy Vayena}, \bibinfo{person}{Dietmar Frey}, \bibinfo{person}{Vince~I Madai}, {and} \bibinfo{person}{Precise4Q Consortium}.} \bibinfo{year}{2020}\natexlab{}.
\newblock \showarticletitle{Explainability for artificial intelligence in healthcare: a multidisciplinary perspective}.
\newblock \bibinfo{journal}{\emph{BMC medical informatics and decision making}}  \bibinfo{volume}{20} (\bibinfo{year}{2020}), \bibinfo{pages}{1--9}.
\newblock


\bibitem[Anderson et~al\mbox{.}(2022)]%
        {anderson@igdb}
\bibfield{author}{\bibinfo{person}{Scott Anderson}, \bibinfo{person}{Loqman Salamatian}, \bibinfo{person}{Zachary~S. Bischof}, \bibinfo{person}{Alberto Dainotti}, {and} \bibinfo{person}{Paul Barford}.} \bibinfo{year}{2022}\natexlab{}.
\newblock \showarticletitle{iGDB: connecting the physical and logical layers of the internet}.
\newblock \bibinfo{journal}{\emph{ACM Meas}} (\bibinfo{year}{2022}), \bibinfo{pages}{433–448}.
\newblock
\showISBNx{9781450392594}
\urldef\tempurl%
\url{https://doi.org/10.1145/3517745.3561443}
\showDOI{\tempurl}


\bibitem[Azab et~al\mbox{.}(2024)]%
        {azab2024network}
\bibfield{author}{\bibinfo{person}{Ahmad Azab}, \bibinfo{person}{Mahmoud Khasawneh}, \bibinfo{person}{Saed Alrabaee}, \bibinfo{person}{Kim-Kwang~Raymond Choo}, {and} \bibinfo{person}{Maysa Sarsour}.} \bibinfo{year}{2024}\natexlab{}.
\newblock \showarticletitle{Network traffic classification: Techniques, datasets, and challenges}.
\newblock \bibinfo{journal}{\emph{Digital Communications and Networks}} \bibinfo{volume}{10}, \bibinfo{number}{3} (\bibinfo{year}{2024}), \bibinfo{pages}{676--692}.
\newblock


\bibitem[Bodria et~al\mbox{.}([n.\,d.])]%
        {bodria2023benchmarking}
\bibfield{author}{\bibinfo{person}{Francesco Bodria}, \bibinfo{person}{Fosca Giannotti}, \bibinfo{person}{Riccardo Guidotti}, \bibinfo{person}{Francesca Naretto}, \bibinfo{person}{Dino Pedreschi}, {and} \bibinfo{person}{Salvatore Rinzivillo}.} \bibinfo{year}{[n.\,d.]}\natexlab{}.
\newblock \showarticletitle{Benchmarking and survey of explanation methods for black box models}.
\newblock \bibinfo{journal}{\emph{Data Mining and Knowledge Discovery}} (\bibinfo{year}{[n.\,d.]}).
\newblock


\bibitem[Dan et~al\mbox{.}(2021)]%
        {dan2021ip}
\bibfield{author}{\bibinfo{person}{Ovidiu Dan}, \bibinfo{person}{Vaibhav Parikh}, {and} \bibinfo{person}{Brian~D Davison}.} \bibinfo{year}{2021}\natexlab{}.
\newblock \showarticletitle{IP geolocation using traceroute location propagation and IP range location interpolation}. In \bibinfo{booktitle}{\emph{Companion Proceedings of the Web Conference 2021}}. \bibinfo{pages}{332--338}.
\newblock


\bibitem[Dan et~al\mbox{.}(2022)]%
        {Dan@traceroute}
\bibfield{author}{\bibinfo{person}{Ovidiu Dan}, \bibinfo{person}{Vaibhav Parikh}, {and} \bibinfo{person}{Brian~D. Davison}.} \bibinfo{year}{2022}\natexlab{}.
\newblock \showarticletitle{IP Geolocation through Geographic Clicks}.
\newblock \bibinfo{journal}{\emph{ACM Trans. Spatial Algorithms Syst.}} \bibinfo{volume}{8}, \bibinfo{number}{1}, Article \bibinfo{articleno}{2} (\bibinfo{date}{March} \bibinfo{year}{2022}), \bibinfo{numpages}{22}~pages.
\newblock
\showISSN{2374-0353}
\urldef\tempurl%
\url{https://doi.org/10.1145/3476774}
\showDOI{\tempurl}


\bibitem[Durairajan et~al\mbox{.}(2015)]%
        {durairajan2015intertubes}
\bibfield{author}{\bibinfo{person}{Ramakrishnan Durairajan}, \bibinfo{person}{Paul Barford}, \bibinfo{person}{Joel Sommers}, {and} \bibinfo{person}{Walter Willinger}.} \bibinfo{year}{2015}\natexlab{}.
\newblock \showarticletitle{{InterTubes: A Study of the US Long‑haul Fiber‑optic Infrastructure}}. In \bibinfo{booktitle}{\emph{Proceedings of the 2015 ACM SIGCOMM Conference}}. \bibinfo{address}{London, United Kingdom}.
\newblock


\bibitem[Elfandi et~al\mbox{.}(2024)]%
        {elfandibootstrapping}
\bibfield{author}{\bibinfo{person}{Abduarraheem Elfandi}, \bibinfo{person}{Hannah Sagalyn}, \bibinfo{person}{Ramakrishan Durairajan}, {and} \bibinfo{person}{Walter Willinger}.} \bibinfo{year}{2024}\natexlab{}.
\newblock \showarticletitle{{Bootstrapping Trust in ML4Nets Solutions with Hybrid Explainability}}. In \bibinfo{booktitle}{\emph{2024 3rd ACM Workshop on Practical Adoption Challenges of ML for Systems}}. ACM.
\newblock


\bibitem[Feng et~al\mbox{.}(2023)]%
        {feng2023degree}
\bibfield{author}{\bibinfo{person}{Qizhang Feng}, \bibinfo{person}{Ninghao Liu}, \bibinfo{person}{Fan Yang}, \bibinfo{person}{Ruixiang Tang}, \bibinfo{person}{Mengnan Du}, {and} \bibinfo{person}{Xia Hu}.} \bibinfo{year}{2023}\natexlab{}.
\newblock \showarticletitle{Degree: Decomposition based explanation for graph neural networks}.
\newblock \bibinfo{journal}{\emph{arXiv preprint arXiv:2305.12895}} (\bibinfo{year}{2023}).
\newblock


\bibitem[Gharaibeh et~al\mbox{.}(2017)]%
        {gharaibeh2017look}
\bibfield{author}{\bibinfo{person}{Manaf Gharaibeh}, \bibinfo{person}{Anant Shah}, \bibinfo{person}{Bradley Huffaker}, \bibinfo{person}{Han Zhang}, \bibinfo{person}{Roya Ensafi}, {and} \bibinfo{person}{Christos Papadopoulos}.} \bibinfo{year}{2017}\natexlab{}.
\newblock \showarticletitle{A look at router geolocation in public and commercial databases}. In \bibinfo{booktitle}{\emph{Proceedings of the 2017 Internet Measurement Conference}}. \bibinfo{pages}{463--469}.
\newblock


\bibitem[Hao et~al\mbox{.}(2021)]%
        {hao2021self}
\bibfield{author}{\bibinfo{person}{Yaru Hao}, \bibinfo{person}{Li Dong}, \bibinfo{person}{Furu Wei}, {and} \bibinfo{person}{Ke Xu}.} \bibinfo{year}{2021}\natexlab{}.
\newblock \showarticletitle{Self-attention attribution: Interpreting information interactions inside transformer}. In \bibinfo{booktitle}{\emph{Proceedings of the AAAI Conference on Artificial Intelligence}}, Vol.~\bibinfo{volume}{35}. \bibinfo{pages}{12963--12971}.
\newblock


\bibitem[Hauff and Houben(2012)]%
        {Hauff@qgis}
\bibfield{author}{\bibinfo{person}{Claudia Hauff} {and} \bibinfo{person}{Geert-Jan Houben}.} \bibinfo{year}{2012}\natexlab{}.
\newblock \showarticletitle{Placing images on the world map: a microblog-based enrichment approach}.
\newblock \bibinfo{journal}{\emph{ACM Meas}} (\bibinfo{year}{2012}), \bibinfo{pages}{691–700}.
\newblock
\showISBNx{9781450314725}
\urldef\tempurl%
\url{https://doi.org/10.1145/2348283.2348376}
\showDOI{\tempurl}


\bibitem[Holterbach et~al\mbox{.}(2015)]%
        {holterbach@ripe}
\bibfield{author}{\bibinfo{person}{Thomas Holterbach}, \bibinfo{person}{Cristel Pelsser}, \bibinfo{person}{Randy Bush}, {and} \bibinfo{person}{Laurent Vanbever}.} \bibinfo{year}{2015}\natexlab{}.
\newblock \showarticletitle{Quantifying Interference between Measurements on the RIPE Atlas Platform}.
\newblock \bibinfo{journal}{\emph{ACM Meas}} (\bibinfo{year}{2015}), \bibinfo{pages}{437–443}.
\newblock
\showISBNx{9781450338486}
\urldef\tempurl%
\url{https://doi.org/10.1145/2815675.2815710}
\showDOI{\tempurl}


\bibitem[Jiang et~al\mbox{.}(2016)]%
        {jiang2016ip}
\bibfield{author}{\bibinfo{person}{Hao Jiang}, \bibinfo{person}{Yaoqing Liu}, {and} \bibinfo{person}{Jeanna~N Matthews}.} \bibinfo{year}{2016}\natexlab{}.
\newblock \showarticletitle{IP geolocation estimation using neural networks with stable landmarks}. In \bibinfo{booktitle}{\emph{2016 IEEE conference on computer communications workshops (INFOCOM WKSHPS)}}. IEEE, \bibinfo{pages}{170--175}.
\newblock


\bibitem[Kaloudi and Li(2020)]%
        {kaloudi2020ai}
\bibfield{author}{\bibinfo{person}{Nektaria Kaloudi} {and} \bibinfo{person}{Jingyue Li}.} \bibinfo{year}{2020}\natexlab{}.
\newblock \showarticletitle{The ai-based cyber threat landscape: A survey}.
\newblock \bibinfo{journal}{\emph{ACM Computing Surveys (CSUR)}} \bibinfo{volume}{53}, \bibinfo{number}{1} (\bibinfo{year}{2020}), \bibinfo{pages}{1--34}.
\newblock


\bibitem[Knofczynski et~al\mbox{.}(2022)]%
        {knofczynski2022arise}
\bibfield{author}{\bibinfo{person}{Jared Knofczynski}, \bibinfo{person}{Ramakrishnan Durairajan}, {and} \bibinfo{person}{Walter Willinger}.} \bibinfo{year}{2022}\natexlab{}.
\newblock \showarticletitle{ARISE: A Multitask Weak Supervision Framework for Network Measurements}.
\newblock \bibinfo{journal}{\emph{IEEE Journal on Selected Areas in Communications}} \bibinfo{volume}{40}, \bibinfo{number}{8} (\bibinfo{year}{2022}).
\newblock


\bibitem[Li et~al\mbox{.}(2017)]%
        {li2017feature}
\bibfield{author}{\bibinfo{person}{Jundong Li}, \bibinfo{person}{Kewei Cheng}, \bibinfo{person}{Suhang Wang}, \bibinfo{person}{Fred Morstatter}, \bibinfo{person}{Robert~P Trevino}, \bibinfo{person}{Jiliang Tang}, {and} \bibinfo{person}{Huan Liu}.} \bibinfo{year}{2017}\natexlab{}.
\newblock \showarticletitle{Feature selection: A data perspective}.
\newblock \bibinfo{journal}{\emph{ACM computing surveys (CSUR)}} \bibinfo{volume}{50}, \bibinfo{number}{6} (\bibinfo{year}{2017}), \bibinfo{pages}{1--45}.
\newblock


\bibitem[Lin et~al\mbox{.}(2022)]%
        {lin2022bert}
\bibfield{author}{\bibinfo{person}{Xinjie Lin}, \bibinfo{person}{Gang Xiong}, \bibinfo{person}{Gaopeng Gou}, \bibinfo{person}{Zhen Li}, \bibinfo{person}{Junzheng Shi}, {and} \bibinfo{person}{Jing Yu}.} \bibinfo{year}{2022}\natexlab{}.
\newblock \showarticletitle{Et-bert: A contextualized datagram representation with pre-training transformers for encrypted traffic classification}. In \bibinfo{booktitle}{\emph{Proceedings of the ACM Web Conference 2022}}. \bibinfo{pages}{633--642}.
\newblock


\bibitem[Liu et~al\mbox{.}(2019)]%
        {liu2019fs}
\bibfield{author}{\bibinfo{person}{Chang Liu}, \bibinfo{person}{Longtao He}, \bibinfo{person}{Gang Xiong}, \bibinfo{person}{Zigang Cao}, {and} \bibinfo{person}{Zhen Li}.} \bibinfo{year}{2019}\natexlab{}.
\newblock \showarticletitle{Fs-net: A flow sequence network for encrypted traffic classification}. In \bibinfo{booktitle}{\emph{IEEE INFOCOM 2019-IEEE Conference On Computer Communications}}. IEEE, \bibinfo{pages}{1171--1179}.
\newblock


\bibitem[Lundberg and Lee(2017a)]%
        {lundberg2017unified}
\bibfield{author}{\bibinfo{person}{Scott~M Lundberg} {and} \bibinfo{person}{Su-In Lee}.} \bibinfo{year}{2017}\natexlab{a}.
\newblock \showarticletitle{A unified approach to interpreting model predictions}.
\newblock \bibinfo{journal}{\emph{Advances in neural information processing systems}}  \bibinfo{volume}{30} (\bibinfo{year}{2017}).
\newblock


\bibitem[Lundberg and Lee(2017b)]%
        {lundberg2017shap}
\bibfield{author}{\bibinfo{person}{Scott~M Lundberg} {and} \bibinfo{person}{Su-In Lee}.} \bibinfo{year}{2017}\natexlab{b}.
\newblock \showarticletitle{A unified approach to interpreting model predictions}.
\newblock \bibinfo{journal}{\emph{Advances in neural information processing systems}}  \bibinfo{volume}{30} (\bibinfo{year}{2017}).
\newblock


\bibitem[Malecki(2002)]%
        {malecki2002economic}
\bibfield{author}{\bibinfo{person}{Edward~J Malecki}.} \bibinfo{year}{2002}\natexlab{}.
\newblock \showarticletitle{The economic geography of the Internet’s infrastructure}.
\newblock \bibinfo{journal}{\emph{Economic geography}} \bibinfo{volume}{78}, \bibinfo{number}{4} (\bibinfo{year}{2002}), \bibinfo{pages}{399--424}.
\newblock


\bibitem[Pujol~Perich et~al\mbox{.}(2021)]%
        {pujol2021netxplain}
\bibfield{author}{\bibinfo{person}{David Pujol~Perich}, \bibinfo{person}{Jos{\'e}~Rafael Su{\'a}rez-Varela~Maci{\'a}}, \bibinfo{person}{Shihan Xiao}, \bibinfo{person}{Bo Wu}, \bibinfo{person}{Alberto Cabellos~Aparicio}, {and} \bibinfo{person}{Pere Barlet~Ros}.} \bibinfo{year}{2021}\natexlab{}.
\newblock \showarticletitle{Netxplain: Real-time explainability of graph neural networks applied to networking}.
\newblock \bibinfo{journal}{\emph{ITU Journal on future and evolving technologies}} \bibinfo{volume}{2}, \bibinfo{number}{4} (\bibinfo{year}{2021}), \bibinfo{pages}{57--66}.
\newblock


\bibitem[Qian et~al\mbox{.}(2024)]%
        {qian2024netbench}
\bibfield{author}{\bibinfo{person}{Chen Qian}, \bibinfo{person}{Xiaochang Li}, \bibinfo{person}{Qineng Wang}, \bibinfo{person}{Gang Zhou}, {and} \bibinfo{person}{Huajie Shao}.} \bibinfo{year}{2024}\natexlab{}.
\newblock \showarticletitle{NetBench: A Large-Scale and Comprehensive Network Traffic Benchmark Dataset for Foundation Models}.
\newblock \bibinfo{journal}{\emph{arXiv preprint arXiv:2403.10319}} (\bibinfo{year}{2024}).
\newblock


\bibitem[Ramanathan and Abdu~Jyothi(2023)]%
        {ram@nautilus}
\bibfield{author}{\bibinfo{person}{Alagappan Ramanathan} {and} \bibinfo{person}{Sangeetha Abdu~Jyothi}.} \bibinfo{year}{2023}\natexlab{}.
\newblock \showarticletitle{Nautilus: A Framework for Cross-Layer Cartography of Submarine Cables and IP Links}.
\newblock \bibinfo{journal}{\emph{Proc. ACM Meas. Anal. Comput. Syst.}} \bibinfo{volume}{7}, \bibinfo{number}{3}, Article \bibinfo{articleno}{46} (\bibinfo{date}{Dec.} \bibinfo{year}{2023}), \bibinfo{numpages}{34}~pages.
\newblock
\urldef\tempurl%
\url{https://doi.org/10.1145/3626777}
\showDOI{\tempurl}


\bibitem[Ramanathan and Abdu~Jyothi(2024)]%
        {Ram@mapping}
\bibfield{author}{\bibinfo{person}{Alagappan Ramanathan} {and} \bibinfo{person}{Sangeetha Abdu~Jyothi}.} \bibinfo{year}{2024}\natexlab{}.
\newblock \showarticletitle{Towards Efficient and Scalable Internet Cross-Layer Mapping}, In \bibinfo{booktitle}{Proceedings of the CoNEXT on Student Workshop 2024} (Los Angeles, CA, USA).
\newblock \bibinfo{journal}{\emph{ACM Meas}}, \bibinfo{pages}{11–12}.
\newblock
\showISBNx{9798400712555}
\urldef\tempurl%
\url{https://doi.org/10.1145/3694812.3699931}
\showDOI{\tempurl}


\bibitem[Ramanathan and Jyothi(2025)]%
        {ramanathan2025leveraging}
\bibfield{author}{\bibinfo{person}{Alagappan Ramanathan} {and} \bibinfo{person}{Sangeetha~Abdu Jyothi}.} \bibinfo{year}{2025}\natexlab{}.
\newblock \showarticletitle{Leveraging Traceroute Inconsistencies to Improve IP Geolocation}.
\newblock \bibinfo{journal}{\emph{arXiv preprint arXiv:2501.15064}} (\bibinfo{year}{2025}).
\newblock


\bibitem[Ribeiro et~al\mbox{.}(2016)]%
        {ribeiro2016should}
\bibfield{author}{\bibinfo{person}{Marco~Tulio Ribeiro}, \bibinfo{person}{Sameer Singh}, {and} \bibinfo{person}{Carlos Guestrin}.} \bibinfo{year}{2016}\natexlab{}.
\newblock \showarticletitle{" Why should i trust you?" Explaining the predictions of any classifier}. In \bibinfo{booktitle}{\emph{Proceedings of the 22nd ACM SIGKDD international conference on knowledge discovery and data mining}}. \bibinfo{pages}{1135--1144}.
\newblock


\bibitem[Robnik-{\v{S}}ikonja and Bohanec(2018)]%
        {robnik2018perturbation}
\bibfield{author}{\bibinfo{person}{Marko Robnik-{\v{S}}ikonja} {and} \bibinfo{person}{Marko Bohanec}.} \bibinfo{year}{2018}\natexlab{}.
\newblock \showarticletitle{Perturbation-based explanations of prediction models}.
\newblock \bibinfo{journal}{\emph{Human and Machine Learning: Visible, Explainable, Trustworthy and Transparent}} (\bibinfo{year}{2018}), \bibinfo{pages}{159--175}.
\newblock


\bibitem[Rusek et~al\mbox{.}(2020)]%
        {rusek2020routenet}
\bibfield{author}{\bibinfo{person}{Krzysztof Rusek}, \bibinfo{person}{Jos{\'e} Su{\'a}rez-Varela}, \bibinfo{person}{Paul Almasan}, \bibinfo{person}{Pere Barlet-Ros}, {and} \bibinfo{person}{Albert Cabellos-Aparicio}.} \bibinfo{year}{2020}\natexlab{}.
\newblock \showarticletitle{RouteNet: Leveraging graph neural networks for network modeling and optimization in SDN}.
\newblock \bibinfo{journal}{\emph{IEEE Journal on Selected Areas in Communications}} \bibinfo{volume}{38}, \bibinfo{number}{10} (\bibinfo{year}{2020}), \bibinfo{pages}{2260--2270}.
\newblock


\bibitem[Sanchez et~al\mbox{.}(2014)]%
        {sanchez2014inter}
\bibfield{author}{\bibinfo{person}{Mario~A Sanchez}, \bibinfo{person}{Fabian~E Bustamante}, \bibinfo{person}{Balachander Krishnamurthy}, \bibinfo{person}{Walter Willinger}, \bibinfo{person}{Georgios Smaragdakis}, {and} \bibinfo{person}{Jeffrey Erman}.} \bibinfo{year}{2014}\natexlab{}.
\newblock \showarticletitle{Inter-domain traffic estimation for the outsider}. In \bibinfo{booktitle}{\emph{Proceedings of the 2014 Conference on Internet Measurement Conference}}. \bibinfo{pages}{1--14}.
\newblock


\bibitem[Selvaraju et~al\mbox{.}(2017)]%
        {selvaraju2017grad}
\bibfield{author}{\bibinfo{person}{Ramprasaath~R Selvaraju}, \bibinfo{person}{Michael Cogswell}, \bibinfo{person}{Abhishek Das}, \bibinfo{person}{Ramakrishna Vedantam}, \bibinfo{person}{Devi Parikh}, {and} \bibinfo{person}{Dhruv Batra}.} \bibinfo{year}{2017}\natexlab{}.
\newblock \showarticletitle{Grad-cam: Visual explanations from deep networks via gradient-based localization}. In \bibinfo{booktitle}{\emph{Proceedings of the IEEE international conference on computer vision}}. \bibinfo{pages}{618--626}.
\newblock


\bibitem[Selvaraju et~al\mbox{.}(2020)]%
        {selvaraju2020grad}
\bibfield{author}{\bibinfo{person}{Ramprasaath~R Selvaraju}, \bibinfo{person}{Michael Cogswell}, \bibinfo{person}{Abhishek Das}, \bibinfo{person}{Ramakrishna Vedantam}, \bibinfo{person}{Devi Parikh}, {and} \bibinfo{person}{Dhruv Batra}.} \bibinfo{year}{2020}\natexlab{}.
\newblock \showarticletitle{Grad-CAM: visual explanations from deep networks via gradient-based localization}.
\newblock \bibinfo{journal}{\emph{International journal of computer vision}}  \bibinfo{volume}{128} (\bibinfo{year}{2020}), \bibinfo{pages}{336--359}.
\newblock


\bibitem[Sengupta et~al\mbox{.}(2022)]%
        {Sengupta@rtt}
\bibfield{author}{\bibinfo{person}{Satadal Sengupta}, \bibinfo{person}{Hyojoon Kim}, {and} \bibinfo{person}{Jennifer Rexford}.} \bibinfo{year}{2022}\natexlab{}.
\newblock \showarticletitle{Continuous in-network round-trip time monitoring}.
\newblock \bibinfo{journal}{\emph{ACM Meas}} (\bibinfo{year}{2022}).
\newblock
\showISBNx{9781450394208}
\urldef\tempurl%
\url{https://doi.org/10.1145/3544216.3544222}
\showDOI{\tempurl}


\bibitem[Simonyan et~al\mbox{.}(2013)]%
        {simonyan2013deep}
\bibfield{author}{\bibinfo{person}{Karen Simonyan}, \bibinfo{person}{Andrea Vedaldi}, {and} \bibinfo{person}{Andrew Zisserman}.} \bibinfo{year}{2013}\natexlab{}.
\newblock \showarticletitle{Deep inside convolutional networks: Visualising image classification models and saliency maps}.
\newblock \bibinfo{journal}{\emph{arXiv preprint arXiv:1312.6034}} (\bibinfo{year}{2013}).
\newblock


\bibitem[Sirinam et~al\mbox{.}(2018)]%
        {sirinam2018deep}
\bibfield{author}{\bibinfo{person}{Payap Sirinam}, \bibinfo{person}{Mohsen Imani}, \bibinfo{person}{Marc Juarez}, {and} \bibinfo{person}{Matthew Wright}.} \bibinfo{year}{2018}\natexlab{}.
\newblock \showarticletitle{Deep fingerprinting: Undermining website fingerprinting defenses with deep learning}. In \bibinfo{booktitle}{\emph{Proceedings of the 2018 ACM SIGSAC conference on computer and communications security}}. \bibinfo{pages}{1928--1943}.
\newblock


\bibitem[Taylor et~al\mbox{.}(2017)]%
        {taylor2017robust}
\bibfield{author}{\bibinfo{person}{Vincent~F Taylor}, \bibinfo{person}{Riccardo Spolaor}, \bibinfo{person}{Mauro Conti}, {and} \bibinfo{person}{Ivan Martinovic}.} \bibinfo{year}{2017}\natexlab{}.
\newblock \showarticletitle{Robust smartphone app identification via encrypted network traffic analysis}.
\newblock \bibinfo{journal}{\emph{IEEE Transactions on Information Forensics and Security}} \bibinfo{volume}{13}, \bibinfo{number}{1} (\bibinfo{year}{2017}), \bibinfo{pages}{63--78}.
\newblock


\bibitem[Thiagarajan et~al\mbox{.}(2025)]%
        {thiagarajan2025aleph}
\bibfield{author}{\bibinfo{person}{Kedar Thiagarajan}, \bibinfo{person}{Esteban Carisimo}, {and} \bibinfo{person}{Fabi{\'a}n~E Bustamante}.} \bibinfo{year}{2025}\natexlab{}.
\newblock \showarticletitle{The Aleph: Decoding Geographic Information from DNS PTR Records Using Large Language Models}.
\newblock \bibinfo{journal}{\emph{Proceedings of the ACM on Networking}} \bibinfo{number}{CoNEXT1} (\bibinfo{year}{2025}).
\newblock


\bibitem[Van~Ede et~al\mbox{.}(2020)]%
        {van2020flowprint}
\bibfield{author}{\bibinfo{person}{Thijs Van~Ede}, \bibinfo{person}{Riccardo Bortolameotti}, \bibinfo{person}{Andrea Continella}, \bibinfo{person}{Jingjing Ren}, \bibinfo{person}{Daniel~J Dubois}, \bibinfo{person}{Martina Lindorfer}, \bibinfo{person}{David Choffnes}, \bibinfo{person}{Maarten Van~Steen}, {and} \bibinfo{person}{Andreas Peter}.} \bibinfo{year}{2020}\natexlab{}.
\newblock \showarticletitle{Flowprint: Semi-supervised mobile-app fingerprinting on encrypted network traffic}. In \bibinfo{booktitle}{\emph{Network and distributed system security symposium (NDSS)}}, Vol.~\bibinfo{volume}{27}.
\newblock


\bibitem[Verma et~al\mbox{.}(2020)]%
        {verma2020counterfactual}
\bibfield{author}{\bibinfo{person}{Sahil Verma}, \bibinfo{person}{John Dickerson}, {and} \bibinfo{person}{Keegan Hines}.} \bibinfo{year}{2020}\natexlab{}.
\newblock \showarticletitle{Counterfactual explanations for machine learning: A review}.
\newblock \bibinfo{journal}{\emph{arXiv preprint arXiv:2010.10596}} (\bibinfo{year}{2020}).
\newblock


\bibitem[Wang et~al\mbox{.}(2025)]%
        {wang2025sigcomm}
\bibfield{author}{\bibinfo{person}{Caleb Wang}, \bibinfo{person}{Ying Zhang}, \bibinfo{person}{Esteban Carisimo}, \bibinfo{person}{Qianli Dong}, \bibinfo{person}{Ram Durairajan}, {and} \bibinfo{person}{Fabián E.~Bustamante}.} \bibinfo{year}{2025}\natexlab{}.
\newblock \showarticletitle{Threading the Ocean: Mapping Digital Routes Across Submarine Cables using Calypso}. In \bibinfo{booktitle}{\emph{ACM SIGCOMM}}.
\newblock


\bibitem[Wang et~al\mbox{.}(2024a)]%
        {wang2024lens}
\bibfield{author}{\bibinfo{person}{Qineng Wang}, \bibinfo{person}{Chen Qian}, \bibinfo{person}{Xiaochang Li}, \bibinfo{person}{Ziyu Yao}, {and} \bibinfo{person}{Huajie Shao}.} \bibinfo{year}{2024}\natexlab{a}.
\newblock \showarticletitle{Lens: A Foundation Model for Network Traffic}.
\newblock \bibinfo{journal}{\emph{arXiv preprint arXiv:2402.03646}} (\bibinfo{year}{2024}).
\newblock


\bibitem[Wang et~al\mbox{.}(2024b)]%
        {wang2024gradient}
\bibfield{author}{\bibinfo{person}{Yongjie Wang}, \bibinfo{person}{Tong Zhang}, \bibinfo{person}{Xu Guo}, {and} \bibinfo{person}{Zhiqi Shen}.} \bibinfo{year}{2024}\natexlab{b}.
\newblock \showarticletitle{Gradient based Feature Attribution in Explainable AI: A Technical Review}.
\newblock \bibinfo{journal}{\emph{arXiv preprint arXiv:2403.10415}} (\bibinfo{year}{2024}).
\newblock


\bibitem[Ying et~al\mbox{.}(2019)]%
        {ying2019gnnexplainer}
\bibfield{author}{\bibinfo{person}{Zhitao Ying}, \bibinfo{person}{Dylan Bourgeois}, \bibinfo{person}{Jiaxuan You}, \bibinfo{person}{Marinka Zitnik}, {and} \bibinfo{person}{Jure Leskovec}.} \bibinfo{year}{2019}\natexlab{}.
\newblock \showarticletitle{Gnnexplainer: Generating explanations for graph neural networks}.
\newblock \bibinfo{journal}{\emph{Advances in neural information processing systems}}  \bibinfo{volume}{32} (\bibinfo{year}{2019}).
\newblock


\bibitem[Zhang et~al\mbox{.}(2023)]%
        {zhang2023tfe}
\bibfield{author}{\bibinfo{person}{Haozhen Zhang}, \bibinfo{person}{Le Yu}, \bibinfo{person}{Xi Xiao}, \bibinfo{person}{Qing Li}, \bibinfo{person}{Francesco Mercaldo}, \bibinfo{person}{Xiapu Luo}, {and} \bibinfo{person}{Qixu Liu}.} \bibinfo{year}{2023}\natexlab{}.
\newblock \showarticletitle{TFE-GNN: A temporal fusion encoder using graph neural networks for fine-grained encrypted traffic classification}. In \bibinfo{booktitle}{\emph{Proceedings of the ACM web conference 2023}}. \bibinfo{pages}{2066--2075}.
\newblock


\bibitem[Zhao et~al\mbox{.}(2024)]%
        {zhao2024explainability}
\bibfield{author}{\bibinfo{person}{Haiyan Zhao}, \bibinfo{person}{Hanjie Chen}, \bibinfo{person}{Fan Yang}, \bibinfo{person}{Ninghao Liu}, \bibinfo{person}{Huiqi Deng}, \bibinfo{person}{Hengyi Cai}, \bibinfo{person}{Shuaiqiang Wang}, \bibinfo{person}{Dawei Yin}, {and} \bibinfo{person}{Mengnan Du}.} \bibinfo{year}{2024}\natexlab{}.
\newblock \showarticletitle{Explainability for large language models: A survey}.
\newblock \bibinfo{journal}{\emph{ACM Transactions on Intelligent Systems and Technology}} \bibinfo{volume}{15}, \bibinfo{number}{2} (\bibinfo{year}{2024}), \bibinfo{pages}{1--38}.
\newblock


\end{thebibliography}
